**Full title**: Non-Destructive, High-Resolution, Chemically Specific, 3D Nanostructure Characterization using Phase-Sensitive EUV Imaging Reflectometry

**Short title**: Phase-Sensitive EUV Imaging Reflectometry

Michael Tanksalvala,[1*] Christina L. Porter,[1] Yuka Esashi,[1*] Bin Wang,[1] Nicholas W. Jenkins,[1] Zhe Zhang,[1] Galen P. Miley,[2] Joshua L. Knobloch,[1] Brendan McBennett,[1] Naoto Horiguchi,[3] Sadegh Yazdi,[4] Jihan Zhou,[5] Matthew N. Jacobs,[1] Charles S. Bevis,[1] Robert M. Karl Jr.,[1] Peter Johnsen,[1] David Ren,[6] Laura Waller,[6] Daniel E. Adams,[1] Seth L. Cousin,[7] Chen-Ting Liao,[1] Jianwei Miao,[5] Michael Gerrity,[1] Henry C. Kapteyn,[1,7] and Margaret M. Murnane[1]

[1] STROBE Science and Technology Center, JILA and Department of Physics, University of Colorado, Boulder, Colorado 80309, USA.
[2] Department of Chemistry, Northwestern University, 2145 Sheridan Road, Evanston, Illinois 60208, USA.
[3] Imec, Kapeldreef 75, 3001 Leuven, Belgium.
[4] Renewable and Sustainable Energy Institute (RASEI), University of Colorado, Boulder, Colorado 80309, USA.
[5] Department of Physics & Astronomy and California NanoSystem Institute, University of California, Los Angeles, California 90095, USA.
[6] Department of Electrical Engineering and Computer Sciences, University of California, Berkeley, CA 94720, USA.
[7] KMLabs Inc., 4775 Walnut St #102, Boulder, Colorado 80301, USA.

[*] Corresponding authors: michael.tanksalvala@colorado.edu, yuka.esashi@colorado.edu

**Abstract (150 words max.)**
Next-generation nano and quantum devices have increasingly complex 3D structure. As the dimensions of these devices shrink to the nanoscale, their performance is often governed by interface quality or precise chemical or dopant composition. Here we present the first phase-sensitive extreme ultraviolet imaging reflectometer. It combines the excellent phase stability of coherent high-harmonic sources, the unique chemical- and phase-sensitivity of extreme ultraviolet reflectometry, and state-of-the-art ptychography imaging algorithms. This tabletop microscope can non-destructively probe surface topography, layer thicknesses, and interface quality, as well as dopant concentrations and profiles. High-fidelity imaging was achieved by implementing variable-angle ptychographic imaging, by using total variation regularization to mitigate noise and artifacts in the reconstructed image, and by using a high-brightness, high-harmonic source with excellent intensity and wavefront stability. We validate our measurements through multiscale, multimodal imaging to show that this technique has unique advantages compared with other techniques based on electron and scanning-probe microscopies.

**MAIN TEXT (maximum of 15,000 words)**

**Introduction**
Although x-ray imaging has been explored for decades, and visible-wavelength microscopy for centuries, it is only recently that the spectral region in between — the extreme ultraviolet (EUV, with wavelengths spanning ~10-100 nm) — has been explored for imaging nanostructures and nanomaterials. This is because high-numerical aperture (NA), high quality optics have not been available in the EUV region of the spectrum. However, with the practical implementation of coherent EUV light sources based on high harmonic generation (HHG), combined with coherent diffraction imaging (CDI)(*1*), EUV imaging has been shown to be competitive in terms of resolution when compared with other light-based imaging techniques (*2–4*). This is important because for synthesis and integration of a host of next-generation materials and nanostructures, new approaches are needed to non-destructively and routinely determine interfacial and layer structure as well as surface morphology, with sensitivity to dopant distributions and material composition. This is becoming more critical as films and devices shrink below 10 nm – where their properties are no longer well-described by bulk macroscopic models and can become almost entirely geometry- or interface-dominated (*5–9*). Moreover, the functional properties of interfaces (i.e. charge, spin, and heat transport) that impact the switching energy of magnetic memory or the coherence time and operating temperature of quantum devices are very difficult to measure, especially in-situ in working devices (*10–12*). As a result, there is a great need for non-destructive, non-contact imaging techniques that can be applied to general samples.

Imaging with EUV light has many unique advantages. It can penetrate materials that are opaque to visible light, making it possible to image buried structures and to extract depth-dependent composition (*13*). When incident at angles between grazing and ~45˚, EUV light has a sufficiently high reflectivity to image most samples (*14–19*). Combined with the fact that the penetration depth of EUV light is sufficiently long to probe interesting structures in most materials, this makes EUV light well-suited for general reflectometry applications. This is in contrast to soft/hard X-ray light at wavelengths <8 nm, which is best-suited for transmission-mode microscopy. Fortunately, high brightness, coherent EUV beams can now routinely be generated via high-harmonic upconversion of intense femtosecond lasers (*20–22*). The low driving laser pulse energies required for HHG—in the 10 µJ to ~mJ range—make it possible to operate at kHz-to-MHz repetition rates that are ideal for applications in imaging and spectroscopy.

When combined with ptychographic CDI (*2, 23–28*), EUV imaging can fill many current characterization gaps. In ptychography, a coherent beam of light is scanned across a sample, and the far-field diffracted intensity is collected from overlapping fields of view. An iterative phase retrieval algorithm is then used to extract quantitative images of the sample's complex transmittance or reflectance from the collected intensity images (*29–31*). Recent advances in CDI are yielding stunning, high-fidelity images, and transforming short-wavelength imaging capabilities (*1, 26, 32–38*). Moreover, by eliminating the need for an image-forming lens, CDI supports diffraction-limited resolution to enable high transverse resolution and axial precision at short wavelengths (*2, 4, 13, 26*). Since ptychographic CDI reconstructs a sample's full complex reflectance, its use with EUV wavelengths is well-suited for phase-sensitive imaging reflectometry applications. In particular, the complex reflectance of coherent EUV light—especially the phase—is exquisitely sensitive to chemical composition, making it possible to determine sample composition uniquely when the incidence angle and/or wavelength of the beam on the sample is scanned (see Fig. 1E and F).

Here we demonstrate the unique advantages of coherent EUV imaging in reflection mode as a general nanoimaging technique. Our previous work has shown that EUV phase-sensitive imaging has promise for measuring many sample parameters – however, in that work, actually determining the sample parameters was a difficult-to-solve, underdetermined problem because it only used a single image of the sample (*13*). Here we show that imaging at many angles of incidence enables us to non-destructively image nanostructures without any special sample preparation or supporting measurements from other metrologies, and with unique 3D compositional specificity that has not heretofore been possible. This fundamentally new technique combines the excellent phase stability of coherent high-harmonic sources with the unique chemical- and phase- sensitivity of extreme ultraviolet reflectometry and state-of-the-art ptychography imaging algorithms. Several aspects were key to implementing high-fidelity phase and amplitude imaging: the ability to correct the glancing-incidence distorted diffraction patterns with very high accuracy, the use of total variation (TV) regularization (*39*) to reduce noise and artifacts in the reconstruction, accurate self-calibration of the reflectometer, and the use of a high brightness, high-harmonic illumination beam that is very stable in intensity, wavelength and wavefront.

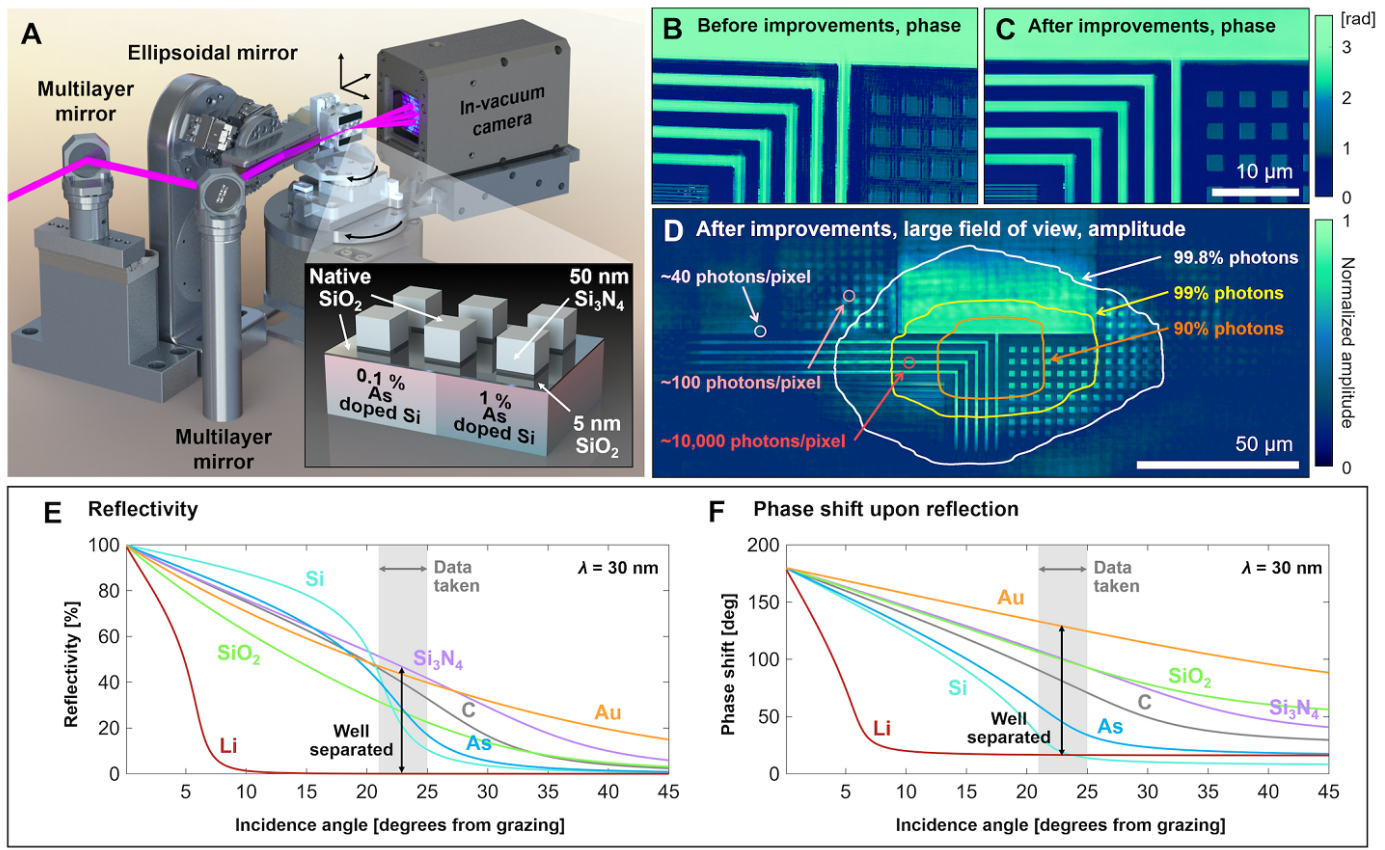

**Fig. 1: Experiment overview and nanostructure imaging.** (**A**) Schematic of the amplitude- and phase- sensitive imaging reflectometer, which produces large-area, spatially- and depth-resolved maps non-destructively. The incidence angle of the illumination is scanned by rotating the sample and detector in a $\theta$-$2\theta$ configuration. The sample can also be scanned in 2D to perform ptychographic coherent diffractive imaging. Inset: Schematic representation of the imaged sample, which has $SiO_2$ + $Si_3N_4$ structures patterned on As-doped regions with higher (~1%) and lower (~0.1%) peak dopant concentration. Native oxide layers ($SiO_2$) are also present. (**B**, **C**) Zoom-in of EUV ptychographic phase reconstructions of the sample, (**B**) before and (**C**) after precise implementation of 3D tilted-plane correction and total variation regularization. (**D**) The entire, wide-field-of-view amplitude reconstruction. Contours with corresponding labels on the right show the regions exposed to certain percentages of total photons that were incident on the sample during a single ptychography dataset. Small circles and corresponding labels on the left indicate the total number of photons that were incident on a pixel at that location over the duration of a single ptychography dataset (light incident at 30° from grazing). (**E**, **F**) Characteristic reflectivity vs. angle curves for several bulk materials at 30 nm wavelength, showing the sensitivity of EUV light to material composition. The phase, measured by our reflectometer but not detected by others, can distinguish between materials even more sensitively than amplitude.

The significance of this advance is that it enables non-destructive, large-area, quantitative, 3D imaging of nanostructures and their chemical makeup, layer thicknesses, interface quality and dopant levels. Moreover, this technique does not require any special sample preparation. The sensitivity that we achieve for some of these parameters is comparable to, or exceeds that of, other techniques that are destructive or contact-based, or that need to average over large unpatterned areas to extract some sample parameters. These include scanning transmission electron microscopy (STEM), secondary ion mass spectrometry (SIMS) and atomic force microscopy (AFM), which were used for correlative imaging in this work. In the future, by harnessing the femtosecond time resolution of EUV HHG beams, the imaging reflectometer can be enhanced further to capture charge, spin and heat transport and link structure to function. The spatial resolution, sensitivity and speed can also be enhanced further by using shorter-wavelength illumination, incidence angles farther from grazing, higher numerical aperture, faster detectors and higher repetition rate drive lasers.

**Experiment setup and procedure for phase-sensitive imaging reflectometry**
To implement phase-sensitive imaging reflectometry, we first record a ptychographic dataset at each incidence angle (see Fig. 1). We used five angles in this initial work — however, with increased data handling capabilities, more angles could be used to solve for more parameters or to lower the uncertainty. Next, we reconstruct an image for each angle using ptychography, giving us quantitative images of the sample's complex reflectance at each angle. These images look similar to one another, but the features change contrast depending on their composition, as well as the wavelength and incidence angle of the illuminating beam. We then segment these images to form reflectance curves for the different sample regions, and using these we reconstruct their depth-dependent compositions using a genetic algorithm. Finally, we combine these into a representation of the sample's topography and composition.

This phase-sensitive imaging reflectometer is illuminated by a coherent high-harmonic beam from a tabletop HHG source (modified prototype KMLabs XUUS4), at wavelengths that can range between 13 and 30 nm. A wavelength of 29.3 nm was used in this initial demonstration to take advantage of the relatively higher reflectivity of the sample at longer wavelengths: the S-polarized reflectivity of passivated Si with a native oxide layer at angles between 21° and 25° from grazing is 3-15%.

The test sample used in this work was custom-fabricated by the Interuniversity Microelectronics Centre (imec). The wafer is a standard 300 mm wafer that used a 65 nm node CMOS mask set, and was doped in different regions with two different doses of As to investigate sensitivity. It had four different types of regions, as shown in the inset of Fig. 1A. The sample is on a silicon substrate that has been selectively doped in some regions to ~1% (atomic percent, corresponding to $5 \times 10^{20}$ atoms/cm$^3$) at an implantation dose of $10^{15}$ atoms/cm$^2$, and then further uniformly doped with a lower dose of $10^{14}$ atoms/cm$^2$. Subsequently, $SiO_2$ and $Si_3N_4$ structures were patterned onto regions of the substrate both with higher and lower doping, while other regions of the substrate were left unpatterned. We will call these four regions higher- and lower-doped substrate and structures, respectively. Native oxide layers ($SiO_2$) are also present on the surface. Further details of the fabrication process can be found in the Methods section.

As shown in Fig. 1A, in our reflectometer both the sample and the EUV CCD camera can be rotated about the focused illumination beam to image the sample at incidence angles ranging from ~10° to 60° from grazing. In this experiment, we recorded a series of five high dynamic range ptychographic images of the sample, one at each incidence angle from 21° to 25° (measured from the sample surface), in 1° increments. In addition, a later scan was recorded with a slightly different field of view (FOV) (shifted by ~20 μm) and angle (30°) for improved visualization.

The phase images from variable-angle ptychographic imaging were first co-registered, and then segmented to measure phase steps within the FOV as a function of the incidence angle: the step between the higher-doped structures and the lower-doped substrate and the phase step between the higher-doped and lower-doped substrate. Pixels within each region were averaged to improve the signal-to-noise ratio (SNR) of the calculated phase steps. Then, to solve for the depth-resolved composition reconstruction, we modeled the sample with the parameters of interest varied about their nominal values (e.g., layer thicknesses, composition, etc.). We then used the Parratt formalism (*40*) to calculate the complex reflectance of our candidate stacks and then iteratively refined parameter estimates with a genetic algorithm that attempted to match the calculated phase steps from the candidate stacks to the measured phase steps. This allowed us to solve for the depth-resolved chemical composition of the sample as well as experimental self-calibration parameters. More information about the genetic algorithm implementation can be found in the SI.

**Results**
**3D nanostructure mapping using phase-sensitive imaging reflectometry**

The first step in the image processing pipeline was to reconstruct a series of ptychographic images of the sample, one at each incidence angle. It is important for these reconstructions to be accurate, since the composition reconstruction is based on the complex reflectances obtained in this step. Figure 1B-D shows the result after several new data preprocessing and image reconstruction procedures were used to increase the fidelity of the images. First, as our microscope is in reflection mode, the diffraction patterns collected at near-grazing angles were interpolated onto a linear spatial frequency grid through a process we call tilted-plane correction (*41*, *42*). We have found that careful implementation of this process (accommodating all three rotation angles of the sample) with accurate parameters is critical for good image fidelity, especially when imaging at a near-grazing angle. Second, we have incorporated TV regularization (*39*) in the reconstruction algorithm to help remove noise and artifacts by favoring solutions with sparse gradients, which is a good assumption for our sample types. Figures 1B and C show reconstructions of the same dataset, with and without these improved procedures, respectively. Note that the image after improvements has sharper structure edges, higher fidelity, and no longer has the skew seen in Fig. 1B. Figure 1D shows the full reconstruction after the improvements. High-fidelity features are reconstructed far beyond the positions of the center of the scanned EUV beam, which roughly corresponds to the FOV shown in Fig. 1C, and even beyond the region encompassing 99% of the photons accumulated over the full scan.

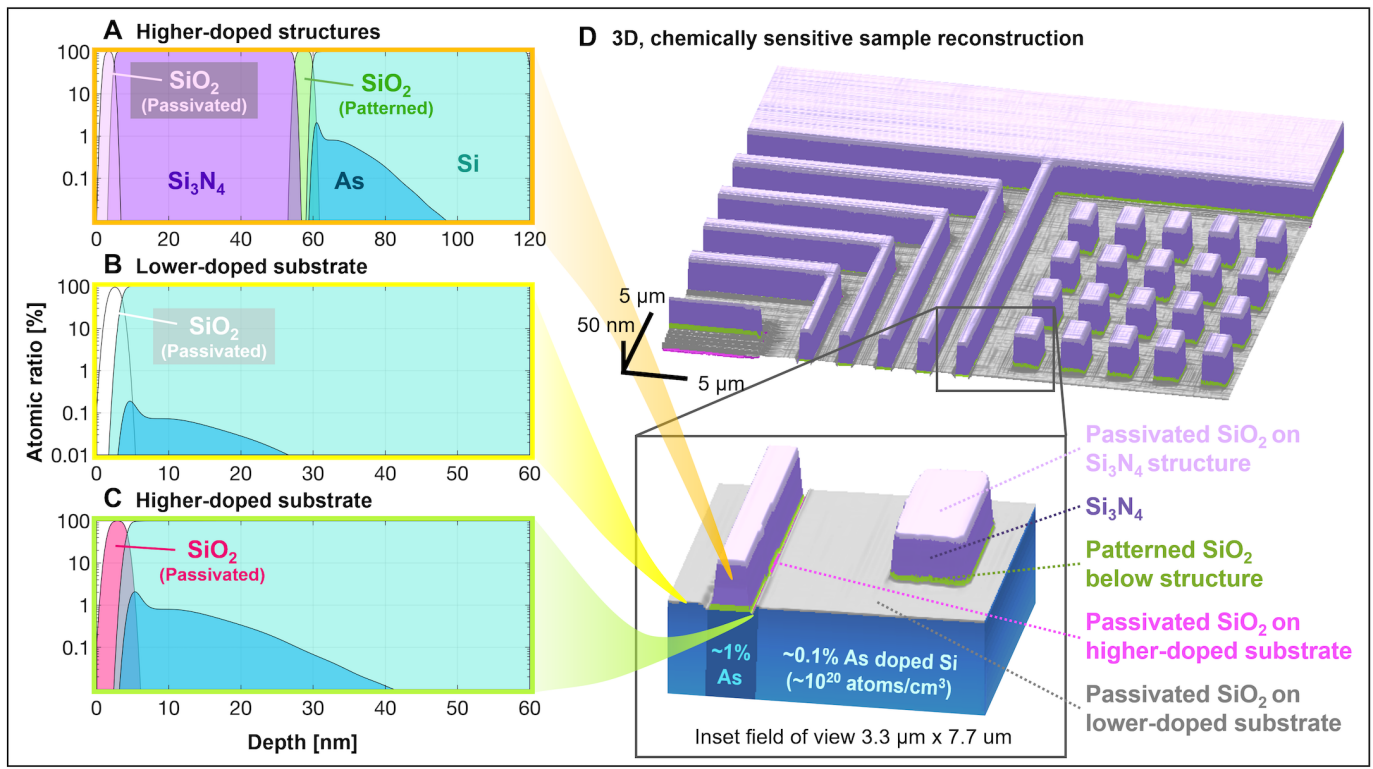

**Fig. 2: Spatially-resolved, composition-sensitive, 3D nanostructure characterization**: Composition vs. depth reconstruction in the (**A**) higher-doped structures, (**B**) lower-doped substrate, and (**C**) higher-doped substrate. The phase-sensitive imaging reflectometer has sensitivity to most parameters within this model (including layer thicknesses and the dopant concentration). Some parameters were determined by correlative imaging (such as surface roughness and interface diffusion). (**D**) Zoom-out and zoom-in (inset) of fully reconstructed sample. This combines the segmented high-fidelity ptychography reconstruction with the material reconstruction from the genetic algorithm, thus showing spatially and depth-resolved maps of material composition, doping, and topography. Different colors correspond to different materials. Notably, different regions of $SiO_2$ are colored uniquely: patterned $SiO_2$ under the structures, passivated $SiO_2$ on higher- and lower-doped substrate, and passivated $SiO_2$ on top of the structures. Also note that we reconstruct the etching adjacent to wide grating lines, shown in magenta in the inset.

Once we had a series of images of the sample's complex reflectance, we segmented the images to calculate the phase steps between different regions of the sample as a function of incidence angle, and using these we solved for the depth-dependent material composition of the sample. Figure 2 highlights the capability of the phase-sensitive imaging reflectometer to non-destructively map, in three dimensions, the chemical composition of general samples. Our reconstructions of the composition vs. depth of different nanopatterned regions (the higher-doped structure, higher- and lower-doped substrate) are shown in Fig. 2A-C. Further parameters that we reconstructed, as well as the calculated sensitivity to those parameters, are listed in Table 1. These can be categorized as parameters related to layer thickness, interface quality and dopant concentration (as well as experiment self-calibration parameters, shown in the Supplementary Materials).

Figure 2D shows the topographic and material map produced by the reflectometer. The different colors in Figure 2 correspond to materials treated differently in the genetic algorithm. For example, there are four colors that each represent a different region of $SiO_2$ (patterned $SiO_2$ under the $Si_3N_4$, passivated $SiO_2$ on top of the $Si_3N_4$, and passivated $SiO_2$ on the higher- and lower-doped substrate). To produce this figure, the complex reconstruction whose phase is shown in Fig. 1C was segmented into the four regions of different compositions using *k*-means clustering, and each segmented region was rendered with the topography and composition solved for by the genetic algorithm (Fig. 3E and 2A-C, respectively). Slight, per-pixel deviations from the resulting height of each region were calculated from the small deviations in the phase from the average phase within the corresponding region.

**Correlative imaging of the nanostructure**
We generated a height map (Fig. 3E) in the same way as the material map (Fig. 2D), to compare with the topography from the AFM measurement. We estimate that the resolution in this image is comparable to the pixel size defined by the detector NA, which is 64 × 172 nm (vertical × horizontal). We have two independent observations supporting this resolution. First, the presence of significant diffraction intensity extending to the vertical and horizontal edges of the detector suggests spatial frequency content that fills that bandwidth. Second, evaluating this resolution from the reconstruction itself, we note that the dopant-related etching seen immediately around the structures in Fig. 3E and F are ~120 nm wide (as measured by AFM). Since this etching has a complex-valued reflectance that is not between the reflectance values of the neighboring regions on either side, this reconstructed feature is not a blurring artifact, and serves the purpose of acting as a resolution-test target. The fact that we see the etching in both directions with enough visibility for the generic *k*-means clustering algorithm to properly segment the image indicates that we have a resolution approaching or surpassing 120 nm in each direction. However, while this is supported by the detector NA in the vertical direction, it is surprising that we see these features even in the horizontal direction, where the etch width is slightly smaller than a single pixel; we expect that this is accompanied by a reduction in visibility of the etching gap. Therefore, from the above two considerations, we conclude that the resolution is comparable to the pixel size defined by the detector NA. The anisotropy in resolution comes from conical diffraction, which stretches the diffraction pattern in the direction of incidence, improving sampling and field of view, but reducing the resolution (*43*).

Because no other single technique could verify all of the parameters that we solved for with phase-sensitive imaging reflectometry, we used several correlative imaging techniques on identical copies of the sample to validate our extracted sample parameters. These techniques included SIMS on an unpatterned sister wafer As-doped to ~1% (Fig. 3D), both high-angle annular dark-field scanning transmission electron microscopy (HAADF-STEM) and energy dispersive X-ray spectroscopy (EDS) (Fig. 3A-C) on a small cross section of the sample, and AFM (Fig. 3F). Note that the first three techniques are destructive, requiring focused ion beam (FIB) milling of the sample, and HAADF-STEM and EDS require special sample preparation. The results of these techniques and phase-sensitive imaging reflectometry are compared in Table 1. The two columns related to phase-sensitive imaging reflectometry outline the promising results of this technique, both when solving for many parameters simultaneously and when solving only for a single parameter (i.e., if the other parameters are known a-priori or measured by other techniques). To avoid over-fitting, we restricted the composition reconstruction to solve only for as many unknowns as we had measured phase steps, so the fixed values in the "simultaneous" column result from the limited number of angles in this initial series of measurements. The values shown in the single-parameter column are intended to give an idea of the order-of-magnitude sensitivity of 30 nm light to each sample parameter. These numbers should not be taken as estimates of the best error bars achievable when solving for a certain parameter: for datasets with more measurements or better-selected wavelength, the error bars could be much better than what is shown, and averaging a greater number of pixels in the segmentation step can also improve these values. In general, we find that phase-sensitive imaging reflectometry has good agreement with other techniques and is sensitive to many sample parameters that are crucial for nanomaterials.

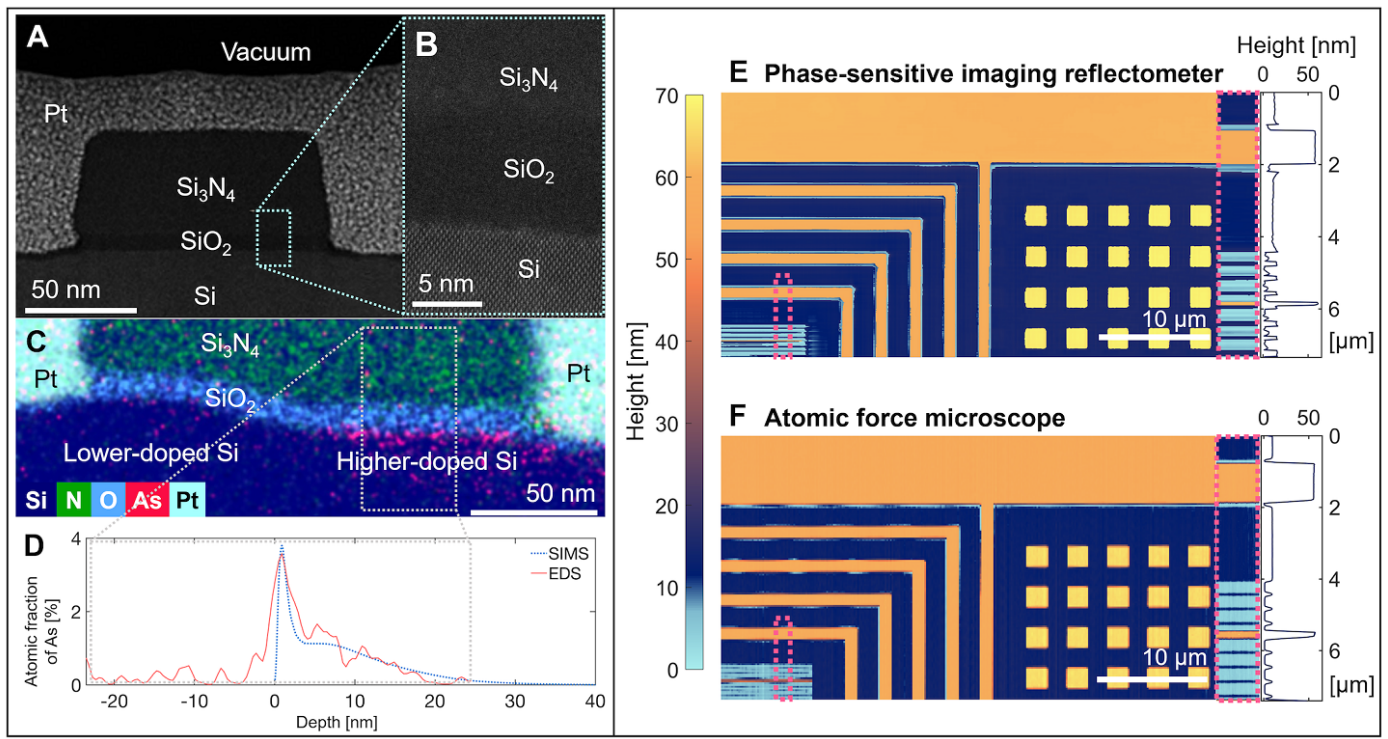

**Figure 3: Correlative imaging with transmission electron microscopy and atomic force microscopy**: (**A**) High-angle annular dark-field scanning transmission electron microscope (HAADF-STEM) image of one of the $Si_3N_4$ structures prepared by focused ion beam (FIB), with (**B**) a zoom-in showing the interfaces between Si, $SiO_2$ and $Si_3N_4$. (**C**) An energy-dispersive X-ray spectroscopy (EDS) image showing a different $Si_3N_4$ structure that is doped to ~1% on the right half and to ~0.1% on the left half. (**D**) The EDS dopant-vs-depth profile that compares well to the curve obtained using SIMS measured on an unpatterned wafer. To increase the signal to noise ratio, the EDS profile was integrated over the area marked by the gray dotted box in C. (**E**) Topography map obtained by combining the ptychographic phase image with the results of the genetic algorithm. The pixel size is 64 nm × 172 nm (vertical × horizontal), and the axial precision is 2 Å. (**F**) Atomic force microscope (AFM) image of the same region. Zoom-in on a region and averaged lineouts of that region are shown on the right.

**Discussion**

The phase-sensitive imaging reflectometry approach introduced and demonstrated here is extremely versatile and general. It is sensitive to almost all of the sample parameters shown in Table 1, and has the capability to reconstruct all of these in a highly spatially- and depth-resolved manner, non-destructively, without contact, and without special sample preparation.

In general, the sensitivity of this technique primarily depends on how many photons can reach

| | Feature | Nominal Value | Phase-Sensitive Imaging Reflectometry | | SIMS* | AFM | EDS / HAADF |
| --- | --- | --- | --- | --- | --- | --- | --- |
| | | | Simultaneous | Single-Parameter Confidence Interval | | | |
| Layer Thickness [nm] | $SiO_2$ on $Si_3N_4$ structure | 0 – 4 | (Set to 3) | ± 0.3 | — | — | 3.0 – 5.0 † |
| | $Si_3N_4$ in structure | 50 | (Set to 50) | Lower bound: 30 | — | — | 41 – 45 |
| | Patterned $SiO_2$ under structure | 5 | (Set to 5) | No sensitivity at 30 nm wavelength | — | — | 6.5 – 7.5 |
| | Structure height | — | 48.2 ± 0.2 | ± 0.02 | — | 45.0 – 45.8 | 48 – 51 |
| | $SiO_2$ on higher-doped substrate | 0 – 4 | 2.7 ± 0.3 | ± < 0.05 | — | — | 2.0 – 4.0 † |
| | $SiO_2$ on lower-doped substrate | 0 – 4 | 2.0 ± 0.3 | ± < 0.05 | — | — | 2.0 – 4.0 † |
| | Dopant-related etch depth | — | 6.09 ± 0.07 | ± 0.02 | — | 7.8 – 8.0 | 5.5 – 7.5 |
| Interface Quality [nm] | Average surface/interface roughness | — | (Set to 0.5) | Upper bound: 0.8 | — | — | 0.5 – 1.0 |
| | Surface roughness on structures | — | (Set to 0.5) | ± 0.2 | — | 0.4 – 0.5 | — † |
| | Surface roughness on lower-doped substrate | — | (Set to 0.5) | ± 0.1 | — | 0.4 – 0.5 | — † |
| | Surface roughness on higher-doped substrate | — | (Set to 0.5) | ± 0.3 | — | 0.4 – 0.5 | — † |
| Dopant | Depth-integrated dose [atoms/$cm^2$] | 1.1e15 | 0.75e15 Upper bound: 5.6e15 | Upper bound: 2.1e15 | 1.05e15 | — | 1.3e15 |
| | Peak concentration [atomic %] | — | (Shape set by SIMS) | Upper bound: 9.3 | 3.8 | — | 3.1 – 4.1 |
| | Gaussian height [atomic %] | — | (Shape set by SIMS) | Upper bound: 3.2 | 1.1 | — | 0.8 – 1.8 |
| Technique Summary | Topography | | Model-Based | | — | Direct | Direct |
| | Composition Information | | Model-Based | | Spectroscopic | — | Spectroscopic |
| | Depth Information | | Model-Based | | Direct | — | Direct |
| | Transverse Spatial Resolution | | Nano-Scale (10-100 nm) | | TOF/nano SIMS: ≥ 100nm | Nano-Scale (10-100 nm) | Atomic Scale (1-100 A) |
| | Field of View | | Meso-to-Micro (10-1000 μm) | | — | Meso (10nm-100 μm) | Atomic-to-Nano (1-1000 nm) |
| | Sample Preparation | | Minimal | | Minimal | Minimal | Versatile, Challenging |
| | Destructive | | Non-Destructive | | Destructive | Contact-Based | Destructive |

**Table 1: Sensitivity of the phase-sensitive imaging reflectometer.** This table compares the reconstructed values of different sample parameters by multiple metrology techniques. The "Nominal Value" column contains the design parameters, as well as expected values from the literature. For phase-sensitive imaging reflectometry, the "Simultaneous" column shows the values simultaneously solved for using the genetic algorithm with the experimental data; only some of the sample parameters were solved for due to the limited number of data points. Were images at more angles available, we expect that we could simultaneously solve accurately for more of these parameters. The "Single-Parameter" column shows the sensitivity to these parameters in a single dimension, measured by how much the fit to the data worsens if an individual parameter is varied around the found solution. This column is a rough estimate of how low the confidence intervals could get with this dataset if we were solving for fewer parameters, and were able to fix the rest using other metrology techniques. The error bars in the phase-sensitive imaging reflectometry columns are given at one standard deviation, while the ranges reported for other techniques, when given, are more loosely defined reasonable ranges given to each measurement. For single-parameter confidence interval calculation, the dopant concentration vs. depth was parametrized as the concatenation of an exponential spike at the surface and a Gaussian extending into the bulk. See Supplementary Materials for complete table.

* The dopant measurements by SIMS were taken on an unpatterned sister wafer. The technique could have made similar measurements on layer thicknesses if there were wafers with the same fabrication steps as this sample, but with much bigger feature sizes (size depends on instrument).
† Variation in the $SiO_2$ thicknesses between phase-sensitive imaging reflectometry (i.e. with phase and amplitude sensitivity) and EDS/HAADF is expected, because the sample had sufficient time to oxidize further between the two measurements. The sample was not prepared to perform surface roughness measurements.

and scatter off of the feature of interest. For instance, 29.3 nm light has a penetration depth of approximately 30 nm in $Si_3N_4$ (*14*), so solving for layer thicknesses of 1-10 nm works extremely well. However, features that lie beneath the 50 nm structures in this sample were more difficult to extract, since at 25° and in reflection, the EUV photons need to pass through $2\times 50/\sin(25°) \approx 250$ nm of material. Using either shorter wavelength ~13 nm light that is more penetrating, or incidence angles farther from grazing, would significantly enhance the ability to detect deeper buried features. Moreover, the sensitivity to interface roughness is proportional to $(\sigma/\lambda)^2 \Delta_n$, where $\sigma$ is the surface roughness, $\lambda$ is the illumination wavelength, and $\Delta_n$ is the difference in the index of refraction of the two layers (see theory section in the Supplementary Materials). Since the surface roughness in this sample is much smaller than 30 nm, the interfaces in this sample represent a challenge to probe using 30 nm light. In the future, by using 13 nm illumination, the sensitivity to interfaces will improve by making $(\sigma/\lambda)^2$ more favorable, while $Si_3N_4$ is ~4x more transparent to 13 nm light (penetration depth of 125 nm) (*14*), so that more photons can reach buried interfaces.

In this demonstration it was shown that this technique can solve for layer thicknesses, dopant levels and experiment self-calibration (see Supplementary Materials). Layer thicknesses, dopant level, and interface quality (to which the single-parameter column in Table 1 shows promising sensitivity) are crucial for proper function of many modern semiconductor, quantum and magnetic devices. With our technique, we were able to set an upper bound on the dopant concentration. We used secondary ion mass spectroscopy (SIMS) to set the shape of the dopant profile and used our data to solve for the dopant concentration. Promisingly, the single-parameter sweep column of Table 1 shows that, even with this preliminary dataset, we have sensitivity to the shape of this curve. Thus, we expect that either with measurements at more incidence angles, or by using 13 nm light, this technique should have sufficient element-sensitivity to resolve this curve's shape natively.

The self-calibration capability allows for more accurate measurement of quantities that are difficult to precisely determine (such as the absolute incidence angle of the illumination on the sample, arising from imperfect sample mounting). By solving for these parameters in the same step that performs the composition reconstruction, we jointly optimize the sample parameters and microscope calibration, and become robust to errors in the system alignment. Furthermore, this system does not suffer as much as other 3D techniques (e.g., tomography) from stringent requirements on physical system alignment that are required to have exactly registered FOVs, since the images are reconstructed independently and can be registered (manually or automatically) after the fact.

We note that the ability to measure dopants in nano-devices is mostly limited to destructive techniques. Auger electron spectroscopy (AES) and SIMS are able to measure dopant levels, but are destructive and typically restricted to unpatterned samples (with a few notable exceptions (*44*, *45*)). On the other hand, TEM based techniques such as off-axis electron holography, EDS and electron energy loss spectroscopy (EELS) can measure dopant concentrations in nanostructures with high spatial resolution, but again are destructive, have challenging sample preparation and are inherently localized techniques with limited FOV (*46–48*).

Model-based scatterometry techniques share some similarities with phase-sensitive imaging reflectometry (*49*), but our technique has unique advantages. Scatterometry-based techniques start from an informed, 3D model of the sample, and solve for sample parameters. While the last step is similar, we can reconstruct the sample without a detailed knowledge of its transverse structure, and require only a 1D model in depth, which requires minimal knowledge of the fabrication steps. In the case that there are uninteresting regions of the sample that are unknown or difficult to model (specs of dust, etc.), our model-free 2D reconstruction can reveal them, and we can avoid them easily by simply omitting those pixels from the depth reconstruction. In contrast, such a feature would often negatively influence the data and the outcome of most scatterometry-based techniques without the user knowing, and even if the user notices the presence of the unexpected feature they would need to turn to an imaging technique for further information.

Finally, we note that a unique advantage of this new technique is its ability to tune the error bars, after the data is already taken, to the sensitivity required to detect a given feature. By selecting larger regions of interest in the image segmentation, the reconstructed composition vs. depth can be improved at the cost of transverse spatial resolution. Of course, there is a limit to the SNR achievable once a dataset is taken, but this ability to tune, almost continuously, the balance of SNR and spatial information to detect a desired feature of interest is remarkable. In high-quality datasets this could allow one to reconstruct a depth profile for every pixel or structure in the FOV, forming a rich 3D map of the sample's composition, topography, and interfaces.

**Conclusion**

We have developed a unique and versatile phase-sensitive imaging reflectometry technique that can non-destructively map the depth-dependent composition of materials, as well as nanostructure layer thicknesses and interface quality, all in a highly spatially-resolved manner. Our results demonstrate that EUV phase-sensitive imaging has exquisite profile sensitivity. By combining the unique strengths of tabletop, coherent, EUV high-harmonic sources with excellent phase stability, with coherent diffraction imaging, we can address imaging science challenges associated with the synthesis and integration of next-generation quantum, semiconductor and spintronic devices and heterostructures - independent of architecture. In the future, it will be possible to enhance the chemical/topographic contrast and the spatial resolution (to <10 nm transverse resolution and <1Å axial precision) by using shorter-wavelength or multi-wavelength illumination, by imaging the sample at multiple in-plane rotational orientations, and/or higher numerical aperture. Moreover, by harnessing the femtosecond time resolution of EUV HHG beams, the imaging reflectometer can be enhanced further to capture charge, spin and heat transport in next-generation devices, and link structure to function. Thus, this work represents a fundamentally new and useful approach for imaging nanostructures and nanomaterials, that has unique advantages compared to complementary techniques such as electron, atomic force and other scanning-probe microscopies.

**Materials and Methods**

**Experimental Design**

To generate the 29.3 nm HHG beam used to illuminate the reflectometer, we focused a femtosecond Ti:sapphire laser (2.1 W, 0.7 mJ, 35 fs, 3 kHz, 0.79 µm) into an argon-filled hollow-core waveguide (150 µm inner diameter, 30 Torr Ar). The resulting flux in the 29.3 nm harmonic was ~$10^{12}$ photons/s at the source, with <1% RMS power stability. This corresponds to $3 \times 10^9$ photons/s at the sample, for the non-optimized beamline used for this initial demonstration experiment. Although ptychographic imaging is robust to noise in the recorded data, it does require that the illumination beam be very stable in intensity, wavelength and wavefront over the course of the scan (*50, 51*). Thus, in our HHG setup, we ensure that we have a very good driving laser stability in both pointing and intensity (0.85% RMS), as well as optimal phase matching conditions.

The residual driving laser light was filtered out using two Si rejector mirrors oriented near Brewster's angle for IR, followed by two 200 nm thick Al filters. A single HHG order was then selected using a pair of SiC/Mg multilayer mirrors. This beam was focused onto the sample using a grazing-incidence ellipsoid, to a spot size of 10 μm × 10 μm at normal incidence for the high-fidelity image, and 21 μm × 21 μm for the imaging reflectometry datasets used for composition reconstruction. Note that a 10 μm diameter focus elongates to as wide as 28 μm at the grazing angles of incidence used in this work. A wavelength of 29.3 nm was used in this initial demonstration to take advantage of the relatively higher reflectivity of the sample at longer wavelengths: the S-polarized reflectivity of passivated Si at angles between 21° and 25° from grazing is 3-15%.

We recorded a series of ptychographic datasets using a 29.3 nm HHG beam as the illumination, with $4 \times 10^7$ photons/s incident on the sample. We collected 5 ptychographic scans on the sample at incidence angles between 21° and 25° (measured from the sample surface), in 1° increments. Each ptychographic scan in the imaging reflectometry dataset contained 301 positions in a Fermat spiral configuration (*52*). Two exposures were collected at each beam position on the sample for high dynamic range (HDR). The lower exposure time at each angle was set 10% shorter than that required to saturate the brightest pixels at that angle, and the longer time was twice that. Before and after data collection at each angle, 150 frames were recorded with the HHG beam pointing directly on the camera (by moving the sample out from the beam and rotating the camera such that the beam is normally incident on the sensor), and background images (with the beam blocked from the system) were recorded for scan and beam data with 0.75 s exposure times. We used the resulting ptychography reconstructions to perform the material reconstruction. These images are shown in the SI, as is information about additional scans not discussed here.

In addition, a later ptychographic dataset was recorded with a slightly different field of view (shifted by ~20 μm) and angle (30°) for improved visualization. This dataset had slightly different parameters. An iris was introduced before the harmonic selecting multilayer mirrors to add structure to the out-of-focus EUV beam. The scan pattern was a rectangular grid instead of a Fermat spiral, and consisted of 424 scan positions (non-HDR), each with an exposure time of 0.1 s. Background data were also recorded with 0.1 s exposure times. We also took beam data before and after the main ptychography scan. These frames were used for the implementation of the modulus enforced probe (MEP) constraint (*4*) on the ptychographic reconstruction of the beam. The resulting image of the sample was not incorporated into the material reconstruction because it was taken with a different system alignment than the imaging reflectometry scan, and would introduce almost as many uncertainties (exact incidence angle, wavelength) as it helped solve for. All of the data was collected with 1×1 binning and a 1 MHz readout rate on a Princeton Instruments (PI) MTE2 CCD (2048 × 2048, 13.5 μm pixels).

**Sample Fabrication**

The fabrication process of the sample studied in this work is illustrated in Fig. S11. First, select regions of the Si wafer were doped with As (5 keV ion implantation) with dose of $10^{15}$ atoms/cm$^2$. Then the photoresist masking the other regions was stripped off, and the entire sample was doped with As with dose of $10^{14}$ atoms/cm$^2$. This created patterns of higher vs lower doped regions on the Si substrate. When the photoresist was stripped, however, it also caused the surface of the unprotected higher-doped regions to be partially removed, creating an unintended dishing of the higher-doped substrate with respect to the lower-doped substrate. The substrate was spike annealed at 1035 °C. Then nominal 5 nm of $SiO_2$ and then 50 nm of $Si_3N_4$ were deposited everywhere on the surface of the sample. Photoresist was then patterned on top of these layers, and the unprotected regions were etched away, creating protruding surface structures. This process resulted in thicker deposited $SiO_2$ underneath the structures, compared to the passivated $SiO_2$ on top of the substrate. There is also passivated $SiO_2$ on top of the $Si_3N_4$ structures.

**Topography and Composition Rendering**

The process of creating the composition-sensitive 3D rendering shown in Fig. 2D involved several processes that ultimately incorporate both the ptychographic reconstruction of the sample and the composition and topography parameters that were solved for in the composition reconstruction.

In general, the phase in ptychographic reconstructions is influenced by both the chemical composition and the surface topography. Within each of the four different regions on our sample (higher- and lower-doped substrate and structures), the mean phase was assumed to come from the average height and the chemical composition, while the variation from the mean was assumed to come from small local topographic variations. Therefore, the ptychographic reconstruction had to be first segmented into the four regions. To do so, first, the phase reconstruction was unwrapped to remove any remaining linear phase ramps in the image so that the phases may be physically interpreted. Then, to accurately render the composition-coded color map for all surfaces (including sidewalls of structures), both the phase and amplitude reconstructions were upsampled 2x using modified Akima interpolation (*53*). In order to produce a map identifying and labeling the four distinct regions of the reconstruction, this complex image was segmented with a *k*-means algorithm, using the amplitude and phase images as the only two channels. To improve robustness of the segmentation, the region containing the fine grating was segmented independently of the remaining part of the sample.

The map of labels produced by the segmentation was then refined through two filtering steps, each performed on separate sections of the reconstruction. The square structures located in the lower right of the image were processed with a binary filter, and the narrow grating located in the lower left was processed with a median filter. Additionally, a very small cluster of clearly mislabeled pixels (likely due to ringing artifact) were manually relabeled. In all, these filtering steps only affected less than 0.7% of the pixels in the image and resulted in label maps that more closely resembled the regions one would identify by eye in the phase and amplitude maps. The remainder of the image did not require any filtering or additional processing.

After the ptychographic reconstruction was segmented into four regions, the average phase of each region was subtracted from the phase map to find the variation from average for every point. This relative phase map was converted to a relative height map by an incidence angle-dependent factor ($h = \Phi/2k \sin(\theta)$, where $h$ is height, $\Phi$ is phase in radians, $k$ is the wave vector and $\theta$ is angles from grazing). The relative height map was then combined with the step heights solved in the composition reconstruction to determine the absolute height of each point in the rendering. After the surface topography was fully determined, independently for each region, composition color coding was performed using the layer thicknesses solved in the composition reconstruction.

**Sensitivity Analysis**

To calculate the confidence interval on each of the fitted parameters, we follow a method that uses the covariance matrix, as described in Press *et al.* (*54*).

To characterize the curvature of the error metric landscape around the found solution, a matrix of double derivatives at that point can be calculated. To an approximation, the elements of this matrix $\boldsymbol{\alpha}$ (referred to as the curvature matrix, or 1/2 of the Hessian matrix) can be expressed using single derivatives:

$$\alpha_{kl} = \sum_{i=1}^{N} \frac{1}{\sigma_i^2} \left[ \frac{d\varphi(\theta_i|a)}{da_k} \frac{d\varphi(\theta_i|a)}{da_l} \right], \tag{4}$$

where $\varphi$ are the measured phase steps, $\varphi(\theta_i|a)$ are the calculated phase steps for the corresponding data point with incidence angle $\theta_i$ and vector of solved-for parameters $\boldsymbol{a}$, and $\sigma_i$ is the standard error of the mean for that data point.

The single derivatives can be approximated numerically using the equation below (higher order methods can be used, but it was found that the discrepancy is negligible):

$$\frac{d\varphi(\theta_i|a)}{da_k} = \frac{\varphi(\theta_i|+\Delta a_k) - \varphi(\theta_i|-\Delta a_k)}{2\Delta a_k}, \tag{5}$$

where $\varphi(\theta_i|+\Delta a_k)$ is the phase step calculated at the found solution but with the $k^{\text{th}}$ parameter displaced by step $+\Delta a_k$.

Once the curvature matrix has been numerically calculated, its inverse known as the covariance matrix $\boldsymbol{C}$ is obtained, and the confidence interval $\delta a_k$ on the $k$th solved parameter is found using its diagonal elements:

$$\boldsymbol{C} = \boldsymbol{\alpha}^{-1} \tag{6}$$

$$\delta a_k = \pm \sqrt{\Delta \chi^2} \sqrt{C_{kk}},$$

where $\Delta \chi^2$ for the desired confidence level can be looked up from a reference table (*54*). The confidence intervals reported in the "Phase-Sensitive Imaging Reflectometry, Simultaneous" columns in Table 1 and Supplementary Table 1 were calculated using this method, for one standard deviation confidence interval.

In addition to the confidence intervals calculated for when parameters were solved simultaneously using a genetic algorithm, we also report the "single-parameter" confidence intervals in a single dimension, measured by how much a parameter can be varied, while the other parameters are fixed, before the error increases by $\Delta\chi^2 = 1$ (54). This column is a rough estimate of how low the confidence intervals could get with this dataset if we were solving for fewer parameters and were able to fix the rest using other metrology techniques.

Using the off-diagonal elements of the covariance matrix and the confidence intervals on the parameters, the correlation coefficient $r$ between the $k^{th}$ and the $l^{th}$ solved-for parameters can also be calculated, using the equation below (54):

$$r_{kl} = \frac{c_{kl}}{\delta a_k \delta a_l}. \tag{7}$$

Fig. 4 shows the correlation coefficient between all the 9 parameters solved for using the genetic algorithm. High magnitude of correlation coefficient means that increase in error due to change in one parameter can be well compensated by change in the other parameter, so it is favorable for correlation coefficients to have low magnitude. In the plot, there are only three parameter pairs that have correlation coefficient > 0.85. Given the strong correlation between the wavelength and the incidence angle offsets, it may have been preferable to fix one of them to the nominal value of zero and only solve for the other, or solve for some factor that combines the two. It is unclear why there is high correlation between $SiO_2$ thickness on higher- and lower-doped substrate, and the carbon deposition rate on wide gratings field of view and the $SiO_2$ thickness on lower-doped substrate; it is possible that this experiment was more sensitive to the difference, or the ratio, of the thicknesses of the layers concerned.

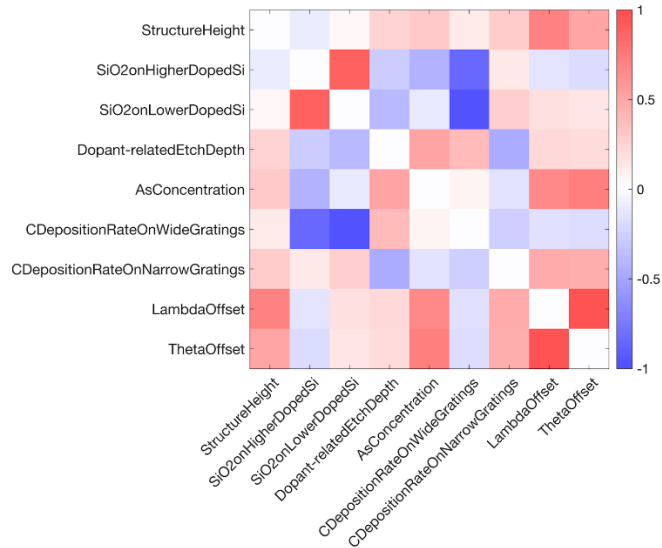

**Fig. 4:** Correlation coefficients between the 9 parameters solved for using the genetic algorithm.

**Acknowledgments:** The authors would like to thank Erik R. Hosler, Kevin Dorney, Jorge N. Hernandez-Charpak, Emma Cating-Subramanian, Stephen Becker, Ian Grooms, David Couch, William Peters, Michaël Hemmer, and Philippe Leray for many helpful discussions. **Funding:** This work was financially supported by NSF STROBE STC (DMR-1548924); DARPA STTR (W31P4Q-17-C-0104); DARPA PULSE (W31P4Q-13-1-0015); Gordon and Betty Moore Foundation's EPiQS Initiative through Grant GBMF; and the following Fellowships: NSF GRFP (1144083), NDSEG Fellowship, China Scholarship Council Joint Doctoral Training Program (No. 201806260040), SRC Fellowship. **Author contributions:** M.T., H.K., C.P. and M.M. conceived this study. M.T., C.P., Y.E., M.G., CT.L., S.C., G.M., H.K. and M.M. designed and/or built the experimental setup. M.T. developed the data collection software. N.H. provided the sample. M.T., Y.E., N.J., C.P., D.A., P.J., D.R., L.W., J.M. and B.M. contributed to the EUV and electron image reconstructions. M.T., C.P., Y.E., N.J., Z.Z., B.W., C.B., S.Y., J.K., J.Z. and R.K. collected the various imaging data and/or analyzed it. All authors contributed to writing the manuscript. **Competing interests:** H.K. and M.M. have a financial interest in KMLabs, that manufactures engineered versions of the prototype ultrafast laser and EUV sources used in this study. M.T., C.P., M.M., H.K., D.A, and E.S. have applied for a patent on this technique. **Data and materials availability:** All data needed to evaluate the conclusions in the paper are present in the paper and/or the Supplementary Materials. Additional data related to this paper may be requested from the authors.

**List of supplementary materials**

    **Sections:**

**Figures:**

# Supplementary Materials: Non-Destructive, High-Resolution, Chemically Specific, 3D Nanostructure Characterization using Phase-Sensitive EUV Imaging Reflectometry

Michael Tanksalvala,[1*] Christina L. Porter,[1] Yuka Esashi,[1*] Bin Wang,[1] Nicholas W. Jenkins,[1] Zhe Zhang,[1] Galen P. Miley,[2] Joshua L. Knobloch,[1] Brendan McBennett,[1] Naoto Horiguchi,[3] Sadegh Yazdi,[4] Jihan Zhou,[5] Matthew N. Jacobs,[1] Charles S. Bevis,[1] Robert M. Karl Jr.,[1] Peter Johnsen,[1] David Ren,[6] Laura Waller,[6] Daniel E. Adams,[1] Seth L. Cousin,[7] Chen-Ting Liao,[1] Jianwei Miao,[5] Michael Gerrity,[1] Henry C. Kapteyn,[1,7] and Margaret M. Murnane[1]

[1] *JILA, STROBE and Department of Physics, University of Colorado, Boulder, Colorado 80309, USA.*
[2] *Department of Chemistry, Northwestern University, 2145 Sheridan Road, Evanston, Illinois 60208, USA.*
[3] *Imec, Kapeldreef 75, 3001 Leuven, Belgium.*
[4] *Renewable and Sustainable Energy Institute (RASEI), University of Colorado, Boulder, Colorado 80309, USA.*
[5] *Department of Physics & Astronomy and California NanoSystem Institute, University of California, Los Angeles, California 90095, USA.*
[6] *Department of Electrical Engineering and Computer Sciences, University of California, Berkeley, CA 94720, USA.*
[7] *KMLabs Inc., 4775 Walnut St #102, Boulder, Colorado 80301, USA.*

[*] *Corresponding authors: michael.tanksalvala@colorado.edu, yuka.esashi@colorado.edu*

## List of Supplementary Materials:

### Sections:

### Figures:

Fig. S9: Intermediate low resolution, wide field of view reconstruction
Fig. S10: Phase step calculation and fit quality
Fig. S11: Schematic of the sample fabrication process
Fig. S12: Angle-dependent reflectometry data collected from the higher-doped and lower-doped substrate

Table S1: Sensitivity of the Phase-Sensitive Imaging Reflectometer

**Phase- and amplitude- sensitive imaging reflectometry: Theoretical sensitivity**

We discuss the sensitivity of this technique to interface quality, layer thickness and composition. We approach this analytically, by examining the expressions for the complex reflectance from rough interfaces and thin films.

First, we investigate the sensitivity to surface roughness. The complex reflectance from a rough surface illuminated near grazing-incidence can be approximated by multiplying the Fresnel coefficient $r$ of the interface (with no roughness) by the Névot-Croce factor $N$ (55, 56). This factor can accurately model either surface roughness or interdiffusion at an interface, as long as the characteristic length scale of the roughness times the perpendicular component of the wavevector is smaller than unity ($\sigma k_z \ll 1$) (56). The factor is defined as:

$$\frac{E'}{E} \cong rN = \begin{cases} \frac{n_1\sin(\theta_1)-n_1\sin(\theta_2)}{n_1\sin(\theta_1)+n_1\sin(\theta_2)} e^{-2n_1 n_2 \left(2\pi\frac{\sigma}{\lambda}\right)^2 \sin(\theta_1)\sin(\theta_2)}, & \text{S} - \text{Pol} \\ \frac{n_1\sin(\theta_2)-n_2\sin(\theta_1)}{n_1\sin(\theta_2)+n_2\sin(\theta_1)} e^{-2n_1 n_2 \left(2\pi\frac{\sigma}{\lambda}\right)^2 \sin(\theta_1)\sin(\theta_2)}, & \text{P} - \text{Pol} \end{cases},$$
(S1)

where $E'/E$ is the ratio of the reflected field $E'$ to the incident field $E$, $r$ is the Fresnel reflectance of the interface, $n_1$ and $n_2$ are the indices above and below the interface, $\sigma$ is the RMS surface roughness, $\lambda$ is the illuminating wavelength, and $\theta_1$ and $\theta_2$ are the propagation angles (from grazing) in the region before and after the interface. Although our complex reflectometer is currently configured to operate with S-polarization, we will show the corresponding expressions for P-polarization for completeness.

For the purposes of this discussion, we define the signal as ($r$ (1 – $N$)), the difference between the reflectance from a perfectly smooth surface and that from a rough surface. In other words, this is how much signal strength we have from the surface roughness. Considering the case where light is impinging from vacuum, we set $n_1 = 1$. Expanding this to first order about $\Delta_n \ll 1$, where $n_2 = 1 + \Delta_n$, and $|1 - N| \ll 1$ gives the following:

$$r(1-N) \approx \begin{cases} -\left(2\frac{\sigma}{\lambda}\right)^2 \Delta_n, & \text{S} - \text{Pol} \\ -\left(2\frac{\sigma}{\lambda}\right)^2 \Delta_n \sin\left(\frac{\pi}{2} - 2\theta_1\right), & \text{P} - \text{Pol} \end{cases}.$$
(S2)

This shows that the signal is proportional to $(\sigma/\lambda)^2$, and increases with $\Delta_n$, the difference in the indices across the interface. Therefore, for a given $\sigma$, using shorter wavelengths can significantly improve the sensitivity to surface roughness. Higher-contrast layers with greater $|\Delta_n|$ will have proportionally higher signal. For reference, using 30 nm light to detect a surface roughness of 0.5 nm on a pure $SiO_2$ substrate gives a signal magnitude of 0.3%, which should be compared to the system noise to determine the feasibility of detection.

In ptychography, we have found that the phase images often have more accurate quantitative information (and are often higher-fidelity) than the accompanying amplitude images. For example, the overall scaling of the object's amplitude reflectance is an unconstrained parameter if the probe is being solved for without further constraints. Therefore, it is useful to determine surface roughness from phase-only measurements. Unfortunately, surface roughness does not affect the phase of the first-order expansion in Eq. S2, since the phaseless coefficients (including $\sigma$) are not detected. Fortunately, expansions that are second-order and higher in $|1 - N|$ do show phase-sensitivity to $\sigma$:

$$r(1-N) \approx \begin{cases} -\left(2\frac{\sigma}{\lambda}\right)^2 \Delta_n \left(1 - \sin^2(\theta_1)\left(2\pi\frac{\sigma}{\lambda}\right)^2\right), & \text{S} - \text{Pol} \\ -\left(2\frac{\sigma}{\lambda}\right)^2 \Delta_n \left(1 - \sin^2(\theta_1)\left(2\pi\frac{\sigma}{\lambda}\right)^2\right) \sin\left(\frac{\pi}{2} - 2\theta_1\right), & \text{P} - \text{Pol} \end{cases}.$$
(S3)

We conclude that, even though phase is often more sensitive to composition and layer thickness than amplitude, surface roughness can be measured more easily by amplitude measurements, if their fidelity and absolute magnitude are sufficiently preserved (though this can be challenging). An intuitive explanation for this is that surface roughness does not change the average path length traveled by photons impinging on the interface, but it does change both the amount of light lost to high-angle scatter and the abruptness of the change in the index of refraction, lowering reflectivity.

Next, in order to examine the sensitivity of the technique to layer thickness or composition, or to the quality of a buried interface, we insert the Névot-Croce factor into the expression for the reflectance of a monolayer, shown here for S-polarization (*57*):

$$\frac{E'}{E} \approx \frac{r_1 + r_2 N_2 e^{-2i\left(2\pi n_2 \frac{T}{\lambda}\right)\sin(\theta_2)}}{1 + r_1 r_2 N_2 e^{-2i\left(2\pi n_2 \frac{T}{\lambda}\right)\sin(\theta_2)}},$$
(S4)

where $r_1$ and $r_2$ are the Fresnel reflection coefficients at the top and the bottom interface (without surface roughness) respectively, $n_2$ and $n_3$ are the refractive indices of the monolayer and the substrate respectively, $N_2$ is the Névot-Croce factor for the interface between the monolayer and the substrate, $T$ is the thickness of the monolayer, and $\theta_2$ is the propagation angle (from grazing) in the monolayer. Here we assume that the interface between the vacuum and the monolayer has no roughness.

We can examine this in a number of informative ways. First, if we assume that the monolayer is sufficiently thick ((imaginary($n_2$)cos($\theta_1$)$T/\lambda$) $\gg$ ($1/2\pi$)) with $\theta_1$ being the propagation angle (from grazing) in the vacuum, then absorption obscures the bottom layer, and the reflectance of the structure approaches that of the top surface. Conversely, if $T, \sigma \ll \lambda$, then the argument of the exponential is small, and a number of approximations can be made. First, we note that the Névot-Croce factor takes the form of an exponential, so Eq. S4 can be written as:

$$\frac{E\prime}{E} \approx \frac{r_1 + r_2 e^{\Phi_{N_2} - i\Phi_{\text{layer}}}}{1 + r_1 r_2 e^{\Phi_{N_2} - i\Phi_{\text{layer}}}}, \tag{S5}$$

where $\Phi_{N_2}$ is the argument of the exponential in the Névot-Croce factor $\left(-2\, n_2\, n_3 \left(2\pi \frac{\sigma}{\lambda}\right)^2 \sin(\theta_2)\sin(\theta_3)\right)$ with $n_3$ and $\theta_3$ the refractive index and propagation angle in the substrate, and $\Phi_{\text{layer}}$ is the argument of the exponential in Eq. S4, which accounts for the phase light accumulates as it travels through the monolayer. If the exponential $e^{\Phi_{N_2} - i\Phi_{\text{layer}}} \approx 1$, then this expression can be approximated:

$$\frac{E\prime}{E} \approx \frac{r_1 + r_2}{1 + r_1 r_2} + \frac{(r_1^2 - 1)r_2}{(1 + r_1 r_2)^2}\left(\Phi_{N_2} - i\Phi_{\text{layer}}\right). \tag{S6}$$

With some algebra, we recognize the first term as the reflectance of the substrate in the absence of the top layer, $r_{13}$:

$$r \approx r_{13} + \frac{(r_1^2 - 1)r_2}{(1 + r_1 r_2)^2}\left(\Phi_{N_2} - i\, \Phi_{\text{layer}}\right). \tag{S7}$$

This is as one might expect: if the layer is very thin, then $\Phi_{\text{layer}}$ approaches zero, and if the roughness correspondingly decreases, $\Phi_{N_2}$ also approaches zero, causing the overall reflectance to match that of the substrate-vacuum interface, $r_{13}$. This term can also be obtained by inserting $T = 0$ and $N_2 = 1$ into Eq. S4. If we expand Eq. S7 to first order in $|1 - N|$ and $\Delta_{n_i}$, we find:

$$\frac{E\prime}{E} \approx \frac{1 - 2\left(\cos^2(\theta_1) + \Delta_{n_3}\tan(\theta_1)\right)}{\left(\cos(\theta_1) + \sin(\theta_1)\right)^2} - \frac{4(\Delta_{n_2} - \Delta_{n_3})\sin(\theta_1)\left(\left(2\pi\frac{\sigma}{\lambda}\right)^2\cos(\theta_1) + i\left(2\pi\frac{T}{\lambda}\right)\right)}{\left(\cos(\theta_1) + \sin(\theta_1)\right)^2}, \tag{S8}$$

where $n_i = 1 + \Delta_{n_i}$. The first term (that dominates the signal if the layer is quite thin) depends only on the incidence angle and the index of the substrate. So, if the layers are quite thin, then the reflectivity is mostly driven by the substrate, which is useful if the goal is to determine the detailed composition of the substrate (e.g., dopant concentration).

The term driving the sensitivity to interface roughness and monolayer thickness is proportional to the contrast between the top and bottom layers $\left(\Delta_{n_3} - \Delta_{n_2}\right)$, indicating that it is easier to interrogate materials that are more dissimilar. 13 nm is a promising wavelength for future experiments: since it is close to the silicon edge, this wavelength range sees a large contrast in indices, and a greatly enhanced sensitivity to silicon-related interfaces or features buried under silicon compounds. Furthermore, the absorption depth argument suggests that moving away from grazing-incidence can also enhance sensitivity to buried features/interfaces, by increasing the number of photons that scatter from the feature and reach the detector.

**Phase- and amplitude- sensitive imaging reflectometry: Numeric simulations**

Figure S1 shows the reflectivity amplitude and phase upon reflection of 30 nm light from different materials. Material sensitivity at this wavelength is indicated by the different shapes of these curves and how spread apart they are from each other, especially as the incidence angle is changed. Both amplitude and phase upon reflection are quite sensitive to a variety of different materials, including light elements such as lithium; however, in general the curves for phase are spread out farther apart. In contrast to most reflectometers, which are sensitive only to the amplitude of the reflectance, our imaging reflectometer that quantitatively measures both phase and amplitude can benefit from the additional information and sensitivity from phase measurements.

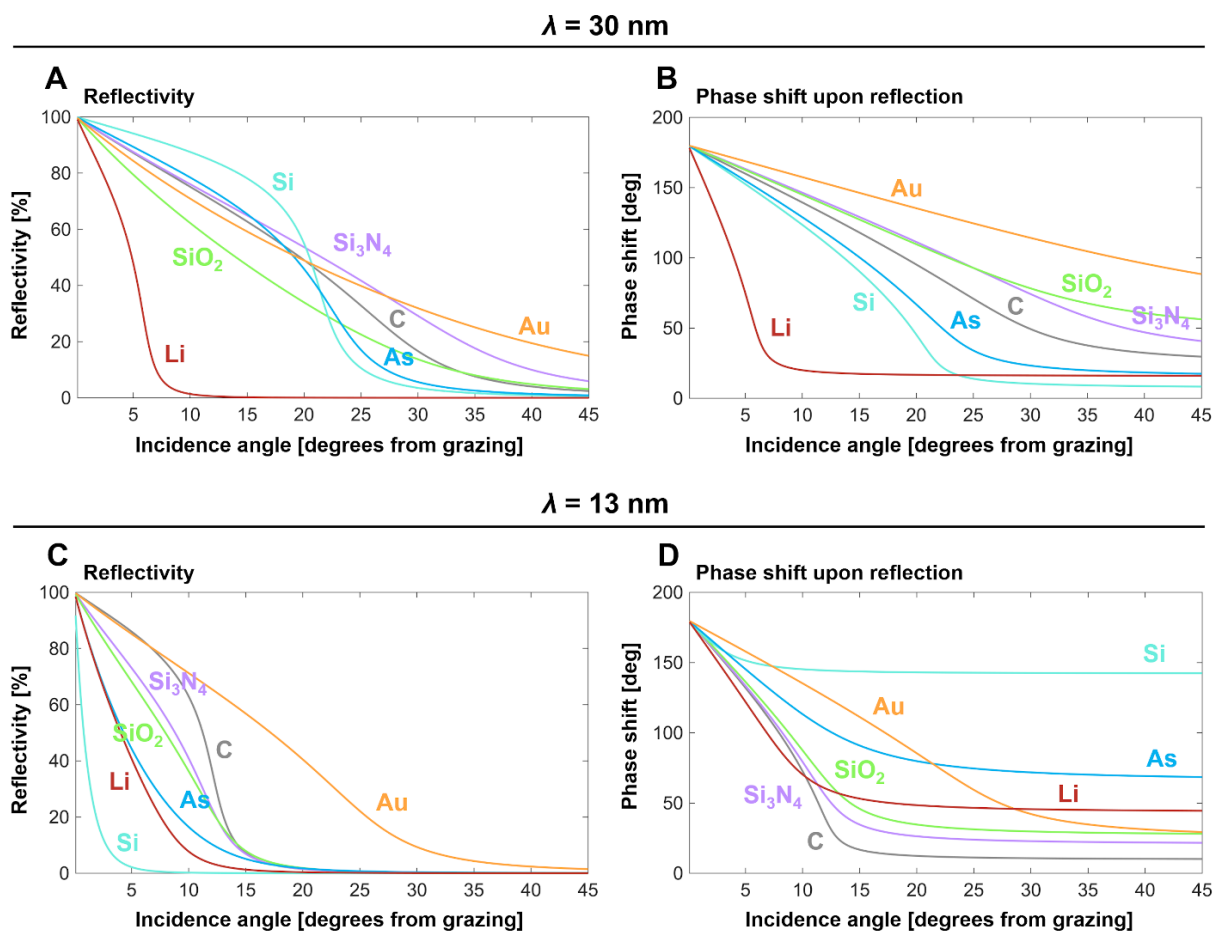

**Fig. S1: Material reflectance vs. incidence angle.** Characteristic reflectivity and phase shift upon reflection vs. angle curves for several materials at 30 nm (**A, B**) and 13 nm (**C, D**) wavelength.

The depth-sensitivity of the technique, as well as the ability to detect buried structures, both are closely related to the absorption depth of the light in the materials comprising the sample. We therefore plot the absorption depth vs. incidence angle for several candidate materials in Fig. S2. Note that this does not take into account reflection of photons off of the top surface before entering the material, so the number of photons that reach a buried structure is lower than what this plot suggests, especially near grazing-incidence.

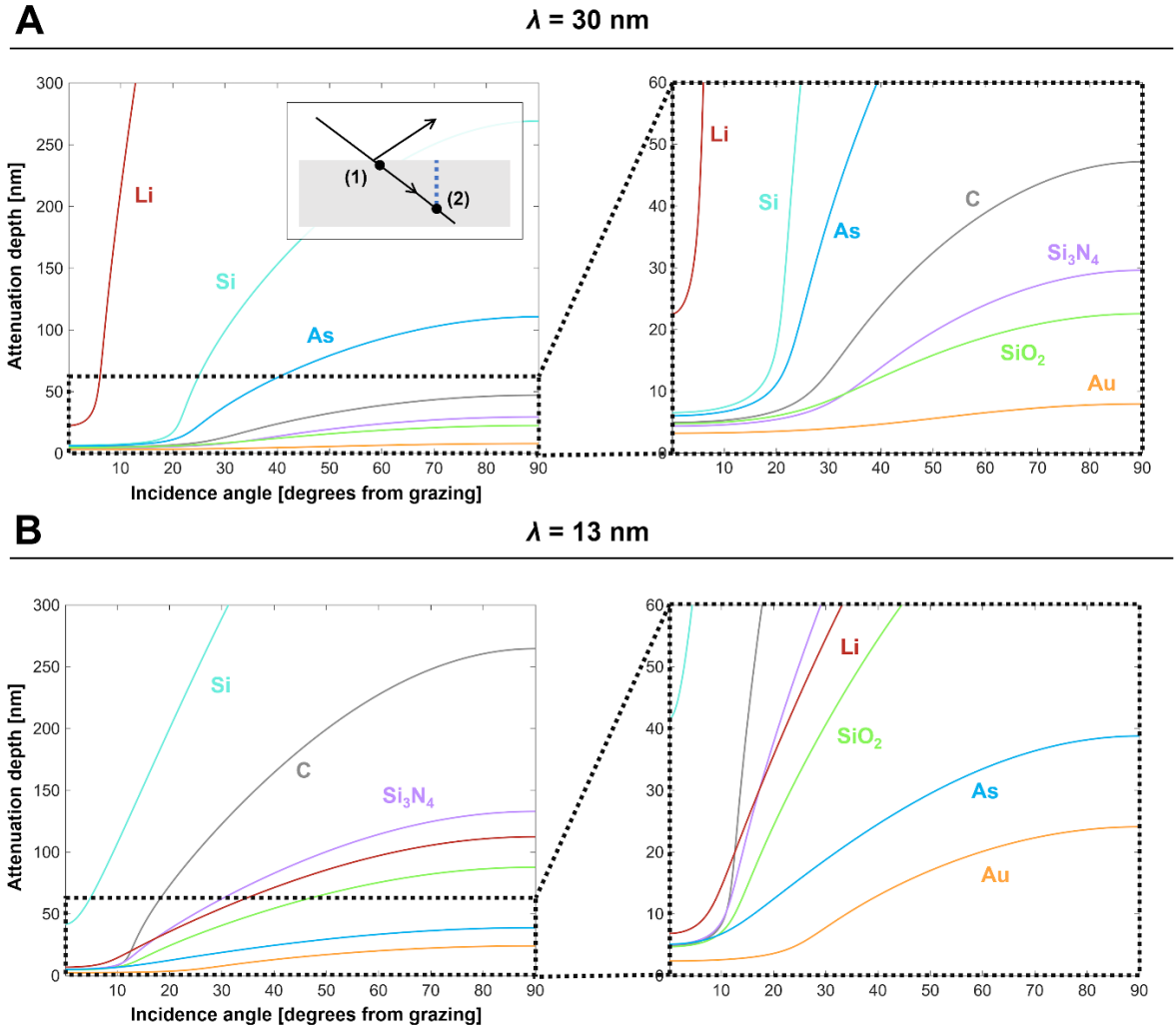

**Fig. S2. Attenuation depths for several materials as a function of incidence angle at 30 nm and 13 nm wavelength.** The attenuation depth is plotted for 30 nm (A) and 13 nm (B) illumination with zoomed-in plots on the right. The attenuation depth is defined as the depth into the material, measured along the surface normal, at which the intensity falls to $1/e$ of its value just inside the surface of the material. That is, given the geometry in the insert, it is the length of the blue dotted line when the ratio of the intensities at (1) and (2) is $1/e$.

To complement the previous theory discussion, we performed numeric simulations with the goal of gaining additional insight into the sensitivity of the technique to a bare substrate's surface roughness, and to several parameters of a monolayer on a substrate, namely interface quality, layer thickness and dopant level. All the simulations are for S-polarization as used in our complex reflectometer.

First, we evaluate the sensitivity of complex reflectometry to the surface roughness of a bare Si substrate. To do so, we model the stack as a graded interface, with the refractive indices changing as an error function from $n = 1$ (vacuum) to the index of the Si substrate. This is an alternative approach to take account of rough interfaces (58) that is different from the Névot-Croce factor that we used in the previous theory section. We then numerically calculate the complex reflectance from the stack using our in-house calculator which is based on the Parratt formalism (40, 59). All the following simulations in this section took the same approach of modeling stacks with graded interfaces.

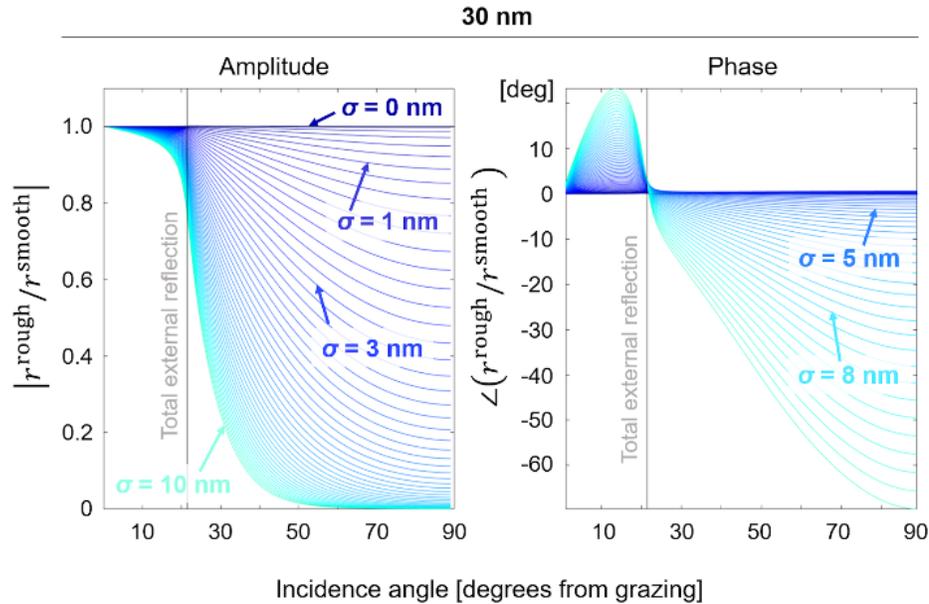

**Fig. S3: Signal from surface roughness on a bare substrate.** Amplitude and phase change of the reflectance ($\lambda = 30$ nm) of a bare Si substrate when surface roughness is introduced, as a function of incidence angle from grazing. The change is expressed as a ratio of the reflectance from a rough surface $r^{rough}$ to the reflectance from a smooth surface $r^{smooth}$. Lines with different colors correspond to different degree of surface roughness $\sigma$, from 0 nm (dark blue) to 10 nm (light blue) in 0.2 nm increments.

In Fig. S3, we plot the amplitude and phase of the complex ratio of the reflectance of 30 nm light from a rough Si surface, to the reflectance of a smooth Si substrate ($r^{rough}/r^{smooth}$), where $r^{smooth}$ is given by the Fresnel equation. The plots are provided for different surface roughnesses, as a function of incidence angle from grazing. For small surface roughness (darker blues), the reflectance is very similar to the reflectance of a smooth substrate, as expected. The reflectance changes substantially as surface roughness approaches 1 nm (amplitude) and 4 nm (phase), and at incidence angles exceeding total external reflection (note that, with complex indices of refraction, the critical angle is complex). This supports our discussion from the previous section, that surface roughness is better detected from amplitude than from phase measurements. While better signal is expected for angles greater than the total external reflection angle in Fig. S3 and many of the following plots, it should be remembered that the absolute reflectivity often drops significantly beyond this angle, meaning that exposure times must increase.

Proceeding in a similar manner to analyze a monolayer, we now simulate a $Si_3N_4$ monolayer on a Si substrate, with a rough interface between the $Si_3N_4$ and Si, as a simple model of the sample we investigated in this paper. We calculate the reflectance of this structure with 30 nm and 13 nm light as the thickness and interface roughness are varied (with zero top surface roughness). Fig. S4 shows the ratio of the resulting reflectance to bulk $Si_3N_4$ ($r^{rough} / r^{Si_3N_4}$), with 30 nm illumination at a 45° incidence angle. The plots are provided for different interface roughnesses, as a function of the monolayer thickness. Note that the curves in these plots are truncated when the surface roughness exceeds half of the thickness of the $Si_3N_4$ layer.

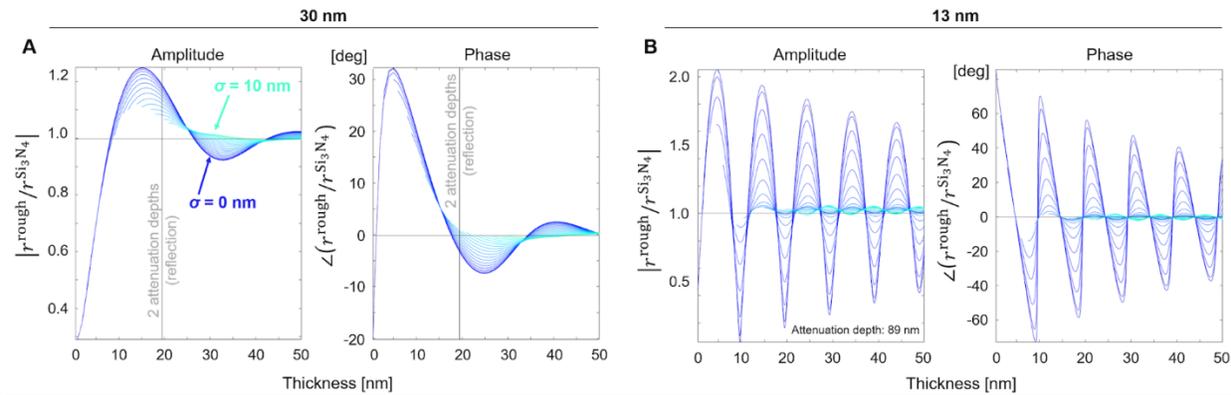

**Fig. S4: Signal vs. thickness of a monolayer with a rough buried interface.** Amplitude and phase change of the reflectance, at (**A**) $\lambda = 30$ nm and (**B**) $\lambda = 13$ nm, of a $Si_3N_4$ monolayer on Si substrate due to interface roughness of the buried interface, as a function of the monolayer thickness. The incidence angle is set at 45°. The change is expressed as a ratio of the reflectance from a monolayer with rough interface $r^{rough}$ to the reflectance from a bulk $Si_3N_4$ with no surface roughness, $r^{Si_3N_4}$. Lines with different colors correspond to different degree of interface roughness $\sigma$, from 0 nm (dark blue) to 10 nm (light blue) in 0.5 nm increments.

As one might expect, when the thickness of the monolayer is much larger than an absorption depth of the illumination wavelength in the monolayer material, the reflectance of the structure approaches that of bulk $Si_3N_4$. Interestingly, increasing the interface roughness (lighter blue) has a similar effect: Since it generally reduces the reflectance of the buried interface, the interference fringes start to disappear and the reflectance approaches that of bulk $Si_3N_4$. At different wavelengths or incidence angles, the particular shape of these curves looks different, but the behavior for very thick samples or very large surface roughness remains largely the same.

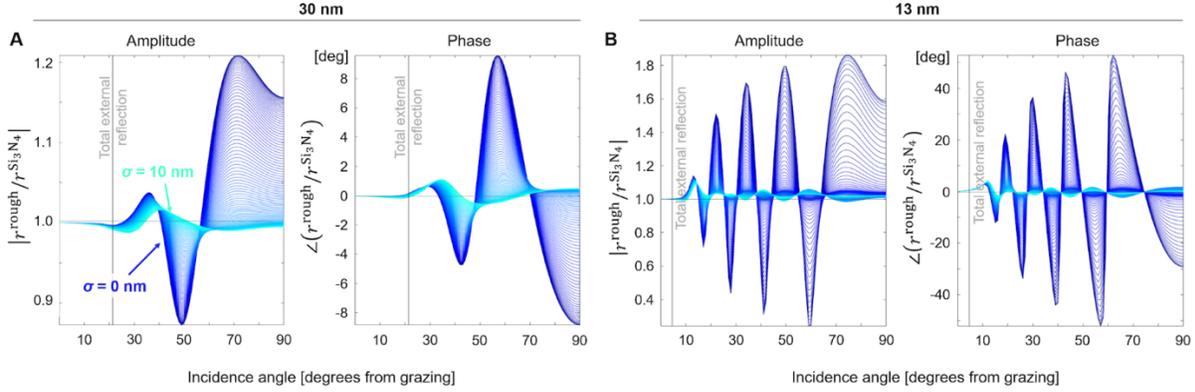

**Fig. S5: Signal vs. incidence angle of a monolayer with a rough buried interface.** Amplitude and phase changes in reflectance, at (**A**) $\lambda = 30$ nm and (**B**) $\lambda = 13$ nm of a 30 nm $Si_3N_4$ monolayer on Si substrate due to interface roughness, as a function of the incidence angle. The change is expressed as a ratio of the reflectance from a monolayer with rough interface $r^{rough}$ to the reflectance from a bulk $Si_3N_4$ with no interface roughness, $r^{Si_3N_4}$. Lines with different colors correspond to different interface roughnesses $\sigma$, from 0 nm (dark blue) to 10 nm (light blue) in 0.1 nm increments.

Fig. S5 shows a similar plot to Fig. S4, except now the thickness is fixed to 30 nm and the incidence angle is varied. At first glance, we see that the signal increases for incidence angles closer to normal incidence. This is because the path length through the material is shorter (meaning less absorption occurs) and more photons are able to reach the second interface and, if reflected, escape the sample and reach the detector. As before, we see that the surface roughness primarily decreases the difference between the reflectance from the monolayer and the bulk reflectance of $Si_3N_4$. Interestingly, the surface roughness affects not only the magnitude but also the shape of this curve. This can be important, because the genetic algorithm used for parameter fitting typically works better if the parameter being solved for changes the shape of the reflectance curve in addition to the overall scaling.

From Figs. S4 and S5, it is understandable that our experiment conducted at wavelength of 30 nm had poor sensitivity to the interface roughness underneath ~50 nm tall $Si_3N_4$ structures, which are significantly more than 2 attenuation depths. The sensitivity would have been higher if the sample had thinner $Si_3N_4$ structures, if the illumination wavelength had longer attenuation depths in $Si_3N_4$, or if the incidence angle was closer to normal incidence, reducing the path length through the structure.

Finally, we look at the sensitivity of the technique to the substrate dopant concentration. To do this, we simulate a layer of $SiO_2$ on an As-doped Si substrate, again as a crude model of the sample used in this paper. For this simulation, we use the dopant profile obtained by SIMS from an unpatterned version of the sample, and we assume that both the top and the buried interfaces have the same roughness (in contrast to the previous two plots, which assumed only the bottom surface had roughness). Fig. S6 shows the ratio of reflectance from the stack with doped substrate to completely undoped substrate ($r^{doped}/r^{undoped}$), with an $SiO_2$ monolayer, for different dopant levels as a function of the incidence angle. Fig. S6A shows a case with 30 nm illumination on a sample with 3 nm thick $SiO_2$ with no surface and interface roughness, and displays relatively weak maximum signal levels of ~6% change in reflectance amplitude and ~1.5° change in phase for dopant concentration of 3x what was measured by SIMS. Fortunately, using 13 nm light on the same sample shown in Fig. S6B has much higher signal levels of ~20% and ~4°. Also note that with 13 nm illumination, there is significant signal level even with 50 nm thick $SiO_2$ and 1 nm of surface and interface roughness, as shown in Fig. S6C. However, it should be noted that the peaks in C are located at points where the reflectivity is very low, so for practical signal levels the regions between the large peaks should be looked at.

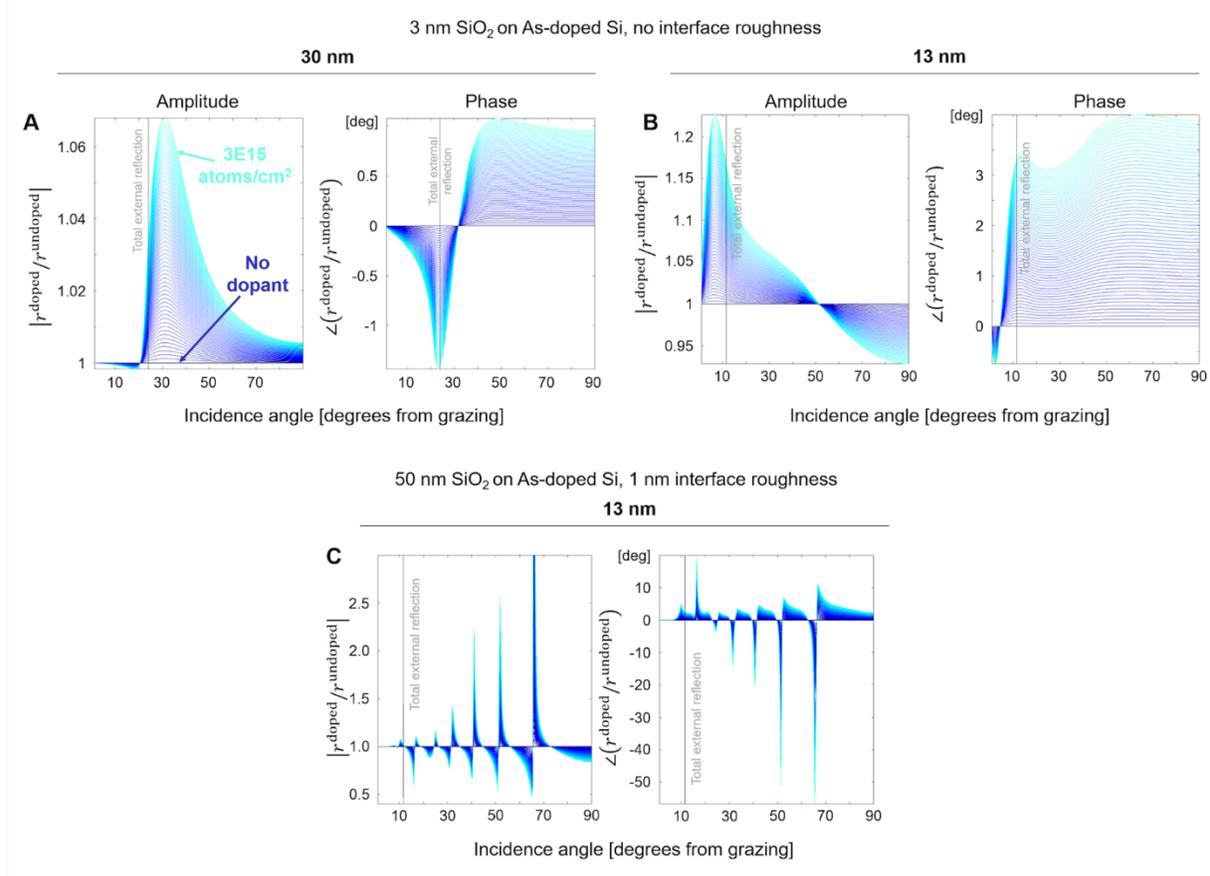

**Fig. S6: Signal from a doped nanostructure.** Amplitude and phase changes in reflectance from a $SiO_2$ monolayer on As-doped Si substrate due to different dopant levels, as a function of the incidence angle. The change is expressed as a ratio of the reflectance from a monolayer on doped substrate $r^{doped}$ to the reflectance from a monolayer on undoped substrate $r^{undoped}$. Lines with different colors correspond to different degree of dopant levels, from 0 atoms/cm$^2$ (dark blue) to ~$3\times10^{15}$ atoms/cm$^2$ (light blue) in ~$5\times10^{13}$ atoms/cm$^2$ increments. For (**A,B**) the $SiO_2$ monolayer has thickness of 3 nm and there is no interface roughness, and the wavelength is (**A**) 30 nm and (**B**) 13 nm. For (**C**), the $SiO_2$ monolayer has thickness of 50 nm, with 1 nm interface roughness, and the wavelength is 13 nm.

The above simulations show that detecting surface and interface roughness and dopant concentration using the parameters used in this experiment (30 nm illumination, 21° -25° from grazing) was a challenge due to low signal levels. Therefore, the fact that we were able to set an upper bound on the dopant concentration, and also demonstrate single-parameter sensitivity to the roughnesses (see Supplementary Table 1) is impressive. In general, for a given sample the signal level can be maximized by optimizing the illumination wavelength and the incidence angle. In addition, if reliable amplitude images are obtained from ptychography, both the amplitude and the phase measurements can be used to further increase sensitivity.

**Future beamline optimization**

In its current state, our instrument can already increase the NA by moving the detector closer to the sample – close enough to support 15nm resolution natively in the vertical direction, when using 13 nm illumination. Moreover, with relatively small modifications, it can be configured to support sub-10 nm vertical resolution. Improving the NA in the other direction can be done in a number of ways; for instance, we can swing the camera independently of the sample to capture an artificially wider NA, or we can image the sample at multiple in-plane rotational orientations. While NA provides a limitation on the resolution, so does the extent of scattered light on the detector (which is optimized by improved material contrast). Many samples can scatter large amounts of light that far, especially grating-type samples relevant to industry. Even samples that are not highly scattering still scatter some photons to the edge of the detector, which can be detected by recording high-dynamic-range diffraction patterns, as is done in this work and was previously published by Helfenstein et al (51).

In the current demonstration, little attempt was made to optimize throughput – for example, a relatively slow CCD camera and software were the primary limitation to the rate of data collection. Thus, we used an HHG beamline with low efficiency – while the flux at the source was $\sim 10^{12}$ photons/s, the flux at the sample was $4 \times 10^7$ photons/s, due to beamline efficiency. (In comparison, more efficient beamlines with higher repetition rate can deliver $> 10^{11}$ photons/s to a sample). As a result, a single ptychographic HHG scan, when the sample is poorly reflective (30°, R~0.012), took in total 15 minutes of HHG exposure, 37 minutes of data saving, and 37 minutes of CCD readout time. Thus, by far the most data acquisition time was limited by the CCD readout and data storage. In total, all exposure time (~5.7 hours), CCD readout time (~5.2 hours), data saving (~5.2 hours) and sample positioning time to allow for settling (~.5 hours) was ~1/2 day. Much of this time could be significantly reduced e.g. the data transfer from the CCD can be enhanced using faster-readout cameras. Or by using existing high efficiency beamlines and taking advantage of the ability of the existing laser to run at higher repetition rates, the on-sample HHG flux can easily be enhanced by >1000x.

**Data collection at other incidence angles**

For this experiment, we collected a total of 18 ptychography datasets in the imaging reflectometry dataset, between 10° and 30° from grazing-incidence. The scans between 21° and 25° were the most important to reconstruct, and we used only these in the main text. Each ptychographic scan in the imaging reflectometry dataset contained 301 positions in a Fermat spiral configuration (*52*). Two exposures were collected at each beam position on the sample for high dynamic range (HDR). The shorter exposure times (set to nearly saturate the camera) ranged between 0.86 seconds at 10° and 15 seconds at 30°, and the longer exposure time was twice that. We randomized the order in which we collected the ptychographic scans at each incidence angle to minimize systematic error such as sample changes due the experiment itself (for instance, hydrocarbon cracking onto the sample). The total sample exposure time (not including beam and background data acquisition time) was 21.9 hours.

**Reconstruction challenges**

This sample proved challenging to reconstruct with high fidelity because it was low-contrast; the materials present were Si, $SiO_2$ and $Si_3N_4$, which have comparable reflectance for 30 nm light. The sample was also three-dimensional with respect to the wavelength, as the structure thicknesses were approximately twice the wavelength. However, in general, the fact that we can examine 3D structures using a series of 2D images, without resorting to a more involved technique like tomography, is a unique strength of imaging reflectometry; since the images can be registered after being individually reconstructed, the requirement for registration of images from different angles is shifted from hardware (difficult) to post-processing (typically easy).

Also, some features on the sample were at or near the system's resolution limit, and there were many regions in the sample that were semi-periodic within one field of view. That is to say, there were many repeated structures within a single beam radius. In this case, we have found that ptychography can have difficulty pinpointing where the end of the repeated structures lies. We note that this is often not an issue in facility-scale experiments, because the beam sizes are often quite small (~40 nm), so this semi-periodicity would have to come from structures that were even smaller (1-4 nm).

**Data pre-processing**

The raw diffraction data was processed prior to running the ptychography reconstruction, with the following steps: first, the ADC offset in the frames were removed by subtracting the mean along each row of overscan columns (200 light-insensitive columns with the same ADC offset as the light-sensitive pixels of the same row) from the data. Then the detector readout noise was removed by use of BM3D (*60*) (imaging reflectometry) or BM4D (*61*) (high-fidelity image) on the diffraction patterns. Further, high-valued pixels due to detection of gamma rays were removed using heuristic gamma ray detection schemes, one based on the ellipticity of nonzero regions of the thresholded diffraction patterns, and the other based on comparing the value of each pixel to its variance across the full dataset. The results of these were visually inspected.

After the noise in the data frames were reduced, any frames that had saturated pixels were discarded in the case of the non-HDR scan, and the two frames taken at each scan position were combined in the case of HDR scans. This was done by normalizing the frames by exposure time, and then replacing values in the longer exposure frames that are half or more of the saturation value, with the corresponding values from the shorter exposure frames.

Finally, we corrected for the three-dimensional tilt of the sample by interpolating the diffraction patterns onto a linear spatial frequency grid through a process called tilted-plane correction (*42*). Multiple combinations of angle tilts around the nominal angles were tested, and the set that made the diffraction pattern the most straight and symmetric was chosen, in order to account for slight misalignment errors of the stages. The diffraction patterns were upsampled when interpolated onto the linear grid to preserve the sampling of the data (*43*). Furthermore, this step divided out the angle-dependent intensity reduction caused by the obliquity factor, the pixel acceptance angle and (for the 30° reconstruction only) the angle-dependent reflectance (45 nm $Si_3N_4$/5 nm $SiO_2$/Si assumed).

**Parallel ptychography code**

We have implemented a ptychographic reconstruction engine (that we call mulptyRAAR) that uses a relaxed, parallel, ptychographic implementation of D. R. Luke's RAAR algorithm (*62*), similar to the algorithm used in SHARP (*63*). It uses the following update:

$$\frac{z}{2}(R_s R_m + I) + (1-z) P_m, \tag{S9}$$

where $R_s$ is the support-constraint reflector $2P_s - I$, $R_m$ is the relaxed modulus constraint reflector $P_m(1+b) - bI$, and $P_m$ is the modulus-constraint projector. $z$ and $b$ are relaxation parameters, and $I$ is the identity operator. Our implementation is written such that beams can have many wavelengths, and moreover the beam and object can each have mutually-incoherent spatial modes (*64, 65*). Our code has available for optional use numerous constraints for the object and probe, related to various sparsity and physicality constraints on the sample and/or probe, which make it extremely useful for general applications. Our code makes heavy use of variable broadcasting in Matlab, for improved speed and memory footprint, and it can furthermore be parallelized over multiple GPSs simultaneously to accommodate datasets too large for a single GPU. In our case, using all eight Tesla P100s on an Nvidia GDX-1 provided a 10x speedup over using all 40 physical cores in the system. Note that the ability to span multiple GPUs is useful for grazing incidence ptychography reconstructions, where high-resolution images are combined with wide fields of view due to the elongated beam profiles at grazing incidence; the use of a single GPU often cannot fulfill the RAM requirements of parallel ptychography.

**Ptychography reconstruction**

We used mulptyRAAR to reconstruct the imaging reflectometry datasets at angles 21° - 25°. We needed to use two incoherent beam modes in the reconstruction, otherwise the reconstruction failed. We believe that this resulted from an instability in our probe that was fixed for the later, high-fidelity dataset. Unfortunately, obtaining quantitatively correct amplitude reconstructions with multimode ptychography is difficult, and for this reason, we only used phase information for our material reconstructions.

After working from a Gaussian-shaped probe guess to get a reconstruction of the nominal 25° scan, the output beam from this scan was used as a guess for the rest of the reconstructions. Although we could have done so, the output object from the 25° reconstruction was not used as a guess for reconstructions at the rest of the incidence angles. The reconstruction at the five angles that were used for the composition reconstruction are shown in Fig. S7.

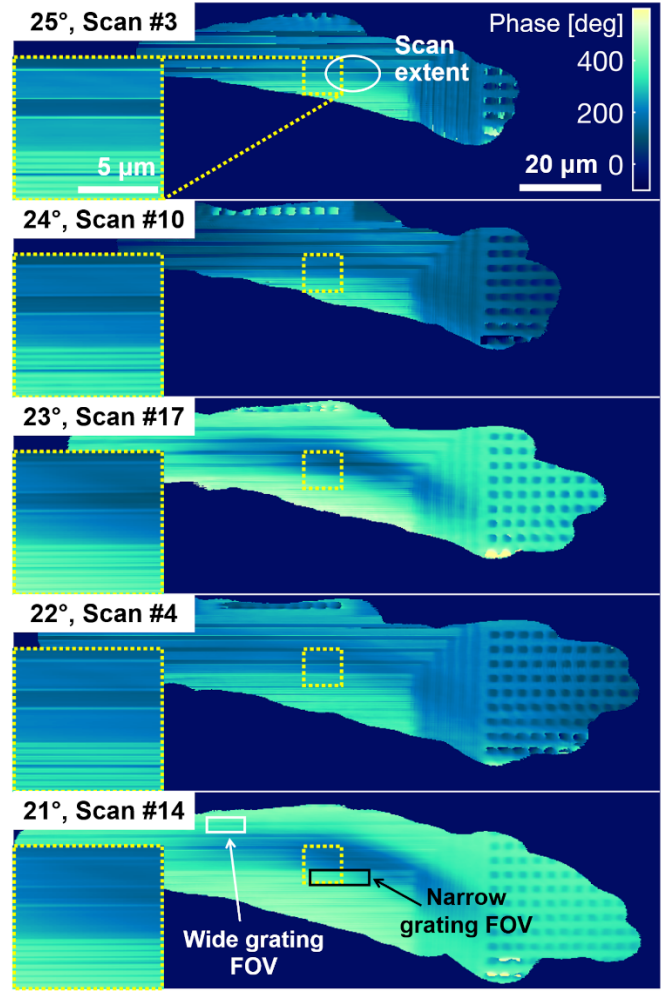

**Fig. S7: Ptychographic reconstructions for phase-sensitive imaging reflectometry.** Five ptychographic reconstructions of the sample at different incidence angles from grazing. The high-fidelity center regions of the reconstructions (indicated by yellow squares) are shown in the insets on the left. The rough scan extent (beam center positions) is marked with a white circle in the top reconstruction, while the two fields of view used to obtain reflectance vs. angle curves are marked with white and black rectangles in the bottom image.

Our procedure for reconstructing the subsequent dataset at 30° was different for a few reasons. Importantly, this dataset did not require multiple probe modes, and as a result, we used a serial implementation of ptychography (based on ePIE) that runs more quickly and with a lower memory footprint but does not incorporate multiple probe modes. We suspect this was due to a few contributing factors, including improved system stability and the added aperture on the beam.

The regularization of the probe was based on previous works. At all iterations, we use RAPTR (*13*) to constrain the power of the probe. Furthermore, we used MEP (*4*) during early iterations to constrain the probe's amplitude on the detector, and a relaxed version of MEP (a support on the detector-space probe) during the remaining iterations.

Regularization of the object was two-fold: we used total variation (TV) regularization, as well as Fourier soft-thresholding, to improve the image fidelity. Every 10 iterations, we used an openly-available TV regularization package (*39*) with regularization parameter $\lambda = 0.05$ on the real and imaginary parts of the object reconstruction to smooth the reconstruction in the high-fidelity region, and used as the object guess the mean of the regularized image and original image. As the high-fidelity region increased, we correspondingly increased the size of the regularized region. In the region with the fine grating, we only performed TV regularization in the horizontal direction.

As one might expect, the TV regularization served to smooth out the noise in the reconstruction. Interestingly, it also made square features on the sample that were previously reconstructing as rounded and blurry into sharp squares. Figure 1 in the main paper shows contributions from both our correct implementation of all three incidence angles in tilted-plane correction, as well as implementation of TV regularization. Both were necessary for the improvement shown: The incidence angles accounted for the skew in the reconstructed image, as well as some of the blurring artifacts, and TV corrected the remaining artifacts.

Furthermore, we used Fourier soft thresholding to regularize the object reconstruction in Fourier-space at every iteration. This involved taking the object's Fourier transform and soft-thresholding its magnitude according to the soft-thresholding operator (wthresh() in Matlab):

$$S(x, \alpha) = \begin{cases} 0 & |x| - \alpha < 0 \\ x - \alpha & \text{else} \end{cases}. \qquad (S10)$$

This promotes sparsity in the object's Fourier transform and, if $\alpha$ is set sufficiently low, it minimizes noise in the reconstruction while leaving the remainder of the reconstruction largely unaffected. This considerably improved the Fourier-space representation of the object as shown in Fig. S8, which we believe have made the real-space representation more accurate, even if it did not produce a visibly noticeable effect.

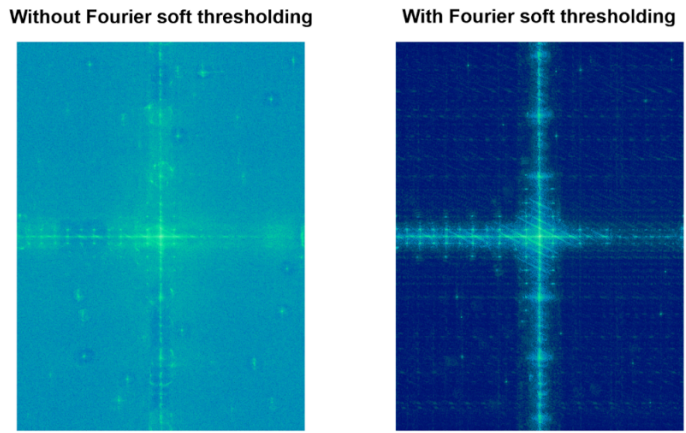

**Fig. S8: Effect of Fourier soft thresholding on the reconstructed object in Fourier space.** Magnitude of the reconstructed object in Fourier space, without (left) and with (right) Fourier soft thresholding. It is clear that the noise floor is considerably lower when Fourier soft thresholding is implemented.

In this 30° reconstruction, we started by reconstructing the dataset with low resolution by cropping the diffraction pattern. This reconstruction has an impressively wide field of view with remarkably little tuning, in a few hundred iterations. After 3000 iterations, it had improved only a little. The full object reconstruction grid at this point is shown in Fig. S9. It is notable that, even though this is blind ptychography with no good starting guess, and the scan grid is only a small portion of the object reconstruction grid as indicated in the figure, the reconstruction extends to nearly the edges of the grid.

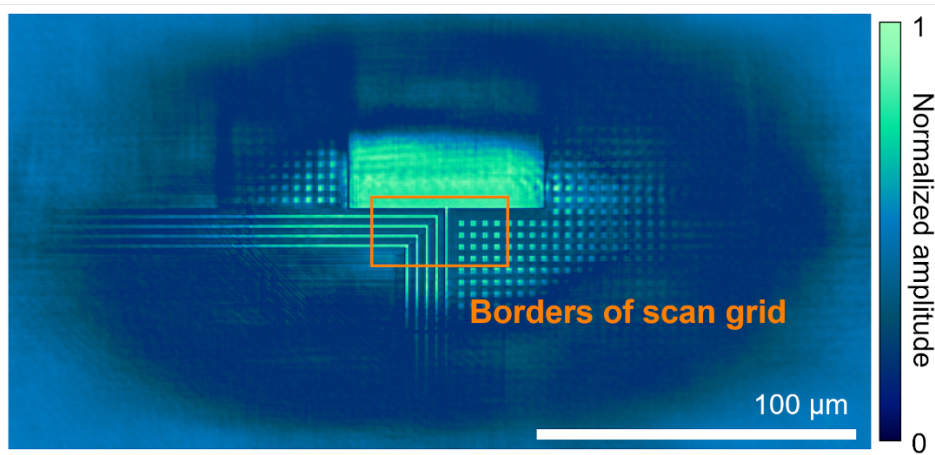

**Fig. S9: Intermediate low resolution, wide field of view reconstruction.** Intermediate amplitude reconstruction at 30° at lower resolution than the final image. Rough borders of the scan grid (scanned beam center) is indicated by the orange rectangle.

We then reconstructed the full dataset without cropping the diffraction patterns, feeding as guesses the object and probe from the low-resolution reconstruction interpolated onto the finer grid. Since this transition merely involved cropping the detector less, the phases of the guessed diffraction patterns in the low-resolution regions remained a good guess. We hoped to retain the wide field of view during the transition from low-resolution to high-resolution reconstructions. To this end, in the first 1000 iterations after this transition, we enforced the amplitude and phase in the modulus constraint in the low-NA regions of the detector. We then ran the modulus constraint as normal for 20,000 iterations. While this number of iterations was probably not necessary, a still-large number of iterations was required to reconstruct the regions that had seen very few photons over the course of the scan. The final reconstruction is shown in Fig. 1 and was used to produce Figures 2D and 3E.

**Phase-sensitive imaging reflectometry reconstructions**

The five reconstructions used in the phase-sensitive imaging reflectometry composition reconstruction are shown in Fig. S7. Even though the scan grid (shown by the white oval in the top frame) does not change size frame to frame, the reconstructions have a wider field of view as the incidence angle moves from 25° to 21°, because the beam's cross section on the sample becomes wider at more grazing angles.

The reconstructed images show discoloration at the center that appears more severe as a function of increasing scan collection number as opposed to incidence angle. This is believed to be due to hydrocarbons in the imaging chamber "cracking" a layer of carbon onto the surface of the sample when the high energy EUV beam interacts with them. This is a known issue in many EUV and electron systems (*66, 67*), and we have since corrected for it by using a PIE Scientific EM-KLEEN plasma cleaner. As will be explained in later sections, we solved for the deposition rate of carbon (nm/hour) to take account of change in reflectance off the sample due to the carbon. It is quite useful that this technique has the ability to accommodate alterations to the sample during data acquisition by solving for the relevant parameters in composition reconstruction.

At each incidence angle, the reconstructed probe showed what appeared to be more than one separate, auxiliary beam on both of the probe scan grids, in addition to the primary one that gave rise to the image reconstructions shown here, requiring reconstruction to be run with multiple probe modes. This meant that the output reflectivity images are not generally quantitative (i.e. they do not always have reflectance between 0 and 100%), because there is an ambiguity within the algorithm about the relative weights between the various probe modes. Thus, the reflectance images that go along with the phase images in Fig. S7 are not bounded between 0 and 100% and therefore could not be used as sources of quantitative absolute reflectance values in composition reconstructions. Although the relative reflectance in these images are probably mostly correct, we also avoided using these values in the composition reconstruction presented here because we have not yet studied the effect of the saturation constraint on relative reflectance in simulation.

We did obtain recognizable images at several incidence angles other than the ones shown in Fig. S7, including the 18°, 19°, 20°, and 29° scans. These images had lower fidelity than the ones presented here, and thus were not deemed reliable enough to use during composition reconstructions. For the more grazing incidence scans, we believe the algorithm struggled to converge in part because our focused beam size was somewhat large—about 15 μm tall by 20 μm wide at normal incidence, projected to 50 μm wide at 25°—for our sample-to-camera distance (48 mm) in this experiment. This gave rise to fairly modest oversampling. Our theoretical oversampling calculator (*43*) predicts that an oval-shaped beam of these dimensions would have minimum and maximum oversampling values on our detector of $(\sigma_{min}, \sigma_{max})$ = (3.6,7.2) at 25°, (3.4,7.2) at 21°, and (2.2,7.2) at 10°. These values are calculated by pretending that the beam had a hard edge at its $1/e^2$ intensity diameter. In fact, the beam had quite an extensive tail (we reconstruct out to about the $1/e^4$ diameter), meaning that realistically ptychography had to deal with very slim oversampling. Our somewhat large beam size was very likely due to a suboptimal alignment of our ellipsoidal focusing optic.

**Image segmentation for producing reflectance vs. angle curves**

The five reconstructed images at angles 21° - 25° were segmented to produce phase step vs. angle curves that were then used in the following composition reconstruction. First, linear phase on the raw phase images, due to the reconstructed probes not lying at the center of their respective probe grids, was removed. This was done by masking the well-reconstructed regions, Fourier transforming these regions, upsampling and then centering the resulting diffraction patterns, then downsampling back to the original pixel size and Fourier transforming back to object space. While most of the linear phase was removed by this method, any residual linear phases were removed at subsequent steps of the processing when deemed necessary, by either fitting 2D planes or 1D lines. Next, the objects were set to zero outside the region illuminated by the beam's $1/e^4$ diameter in at least 4 positions during the ptychography scan, since regions of the object grid that were never sufficiently illuminated should not be considered.

Then the phase images were unwrapped, removing discontinuities in the images due to the reconstructed phase being bounded between 0 and $2\pi$ radians. This was done using an in-house algorithm, which takes as an input manually drawn masks where the image should increase or decrease by $2\pi$ radians. The algorithm then scans the pixels in the masks and increases/decreases each pixel by $2\pi$ when it would result in a lower standard deviation in a window around that pixel.

The unwrapped images were registered to each other using the upsampled cross-correlation method of Guizar-Sicairos *et al*. (*68*). Within the registered images, two regions of interest were chosen – one containing wide, higher-doped structure gratings on lower-doped substrate, and the other containing narrow, higher-doped substrate gratings on lower-doped substrate. Care was taken to avoid regions of the field of view that showed the most carbon contamination due to the EUV photons while still choosing areas with relatively high fidelity in all of the frames, since reconstructing the carbon profile was not the main goal here. The regions were averaged in the horizontal direction, parallel to the grating lines in which there was little change in the phase values, to create average grating lineouts. Then the lineouts were thresholded to measure the average local phase values: in the wide grating field of view, higher-doped structure and lower-doped substrate phase values were calculated, while in the narrow grating field of view, higher- and lower-doped substrate phase values were calculated. Examples of lineouts and the selected points are shown in Fig. S10A and B. Some manual adjustment of the selected points after thresholding was necessary to make sure that irrelevant points were not selected. For example, in the wide grating field of view, there are narrow channels on either side of the structure gratings caused from a dopant-related etch (discussed in the Sample Fabrication section of the main text). Some pixels on the edges of the wide grating are also affected by ringing artifacts. In the narrow grating field of view, there are a few structure grating lines that must be avoided.

Note that this highlights a major benefit of this technique: by splitting up the 2D ptychographic image reconstruction step from the third dimension (depth) reconstruction, "local issues" can be avoided such as sample contamination, unexpected phenomena like the etching channels, and image reconstruction artifacts like ringing near edges. In addition, for high quality reconstructions, only averaging of a small subset of reconstructed image is necessary to calculate average phase steps with small enough error bars. This means that the composition reconstruction can be conducted on a less-spatially averaged region, and ultimately this enables the technique to reconstruct a depth profile for every pixel or structure in the field of view, forming a rich 3D map of the sample's composition, topography, and interfaces.

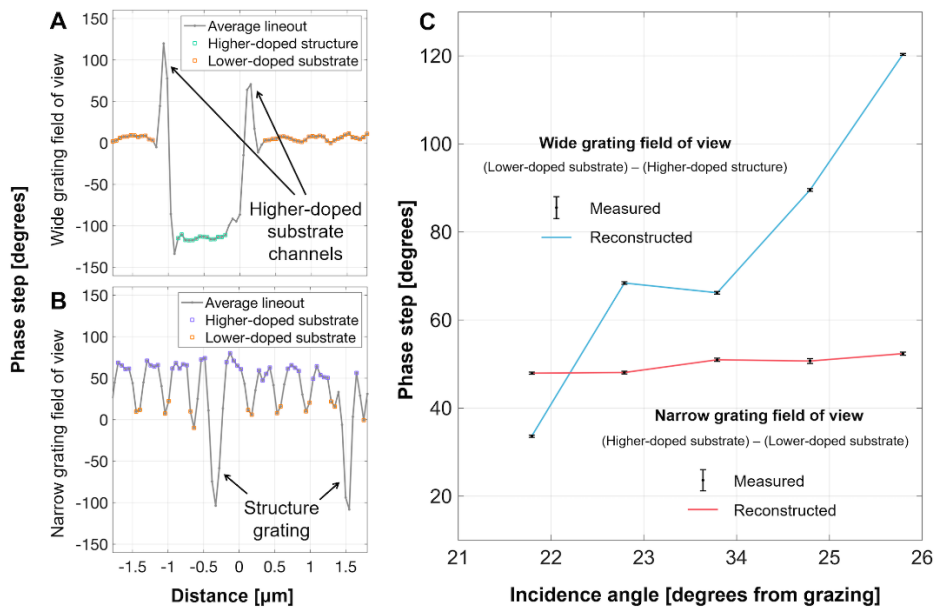

**Fig. S10: Phase step calculation and fit quality.** Lineouts of (**A**) wide- and (**B**) narrow-grating fields of view from the phase of the ptychography image from 25° incidence angle. The colored points are the points used to calculate the phase steps. Certain features, including higher-doped substrate channels in the wide grating field of view and structure lines in the narrow grating field of view, were manually excluded from the selection. Shown in (**C**) are experimentally measured phase steps and the fit by genetic algorithm. Phase steps measured in the wide grating field of view (top curve) and narrow grating field of view (bottom curve) are shown as black points. The error bars indicate the standard error of the mean of the pixels used in the segmentation, and their sizes are between 0.2° and 0.6°. The reconstructed curves for the two fields of view are shown in blue and red curves respectively.

The average phase steps in these areas as a function of incidence angle are shown by the black datapoints in Fig. S10C. The error bars show the standard error of the mean of all the pixels from the reconstruction that were averaged to create that data point.

**Composition reconstruction parameter selection**

We selected 9 parameters to solve for from our phase vs. incidence angle curves. Two of the parameters were related to the surface topography, two to layer thicknesses, one to the As dopant levels, and the rest to system calibration parameters.

We solved for two different profile heights on the sample: the "structure step" between the lower-doped substrate and the higher-doped structures (in the wide grating field of view), and the "dopant-related etch step" from higher-doped to lower-doped substrate (in the narrow grating field of view) that were created due to the removal of photoresist that covered the lower-doped parts of the substrate, as shown in Fig. S11.

We solved for different thicknesses of $SiO_2$ on top of the higher- and lower-doped substrate. This was justified by the fact that solving for two thicknesses produced significantly better fits. In addition, we also collected angle-dependent absolute reflectivity measurements off the higher- and lower-doped substrate, as shown in Fig. S12. The two curves were sufficiently different to imply that these reflectivity variations could not have resulted from the different dopant levels alone, according to our simulations. We also believe that the absolute reflectivity difference could not have come from different surface roughness between the two regions, because AFM measurements on these two regions of the sample only showed small (<0.02 nm) difference in surface roughness. As for other layer thicknesses within the sample, we did not solve for the $Si_3N_4$ thickness in the structures and the underlying $SiO_2$ thickness; they were fixed to the nominal design parameters of 50 nm and 5 nm respectively. A very small amount of 30 nm light would have reached the bottom $Si_3N_4/SiO_2$ interface underneath the structure, given the large thickness of $Si_3N_4$ and absorption length of 30 nm light in these materials. We also did not solve for the thickness of the $SiO_2$ on top of the $Si_3N_4$, and this value was set to 3 nm—while preliminary analysis showed that our dataset was sensitive to this thickness, when this parameter was solved for in the composition reconstruction and its sensitivity calculated, the sensitivity was significantly lower than that of other parameters.

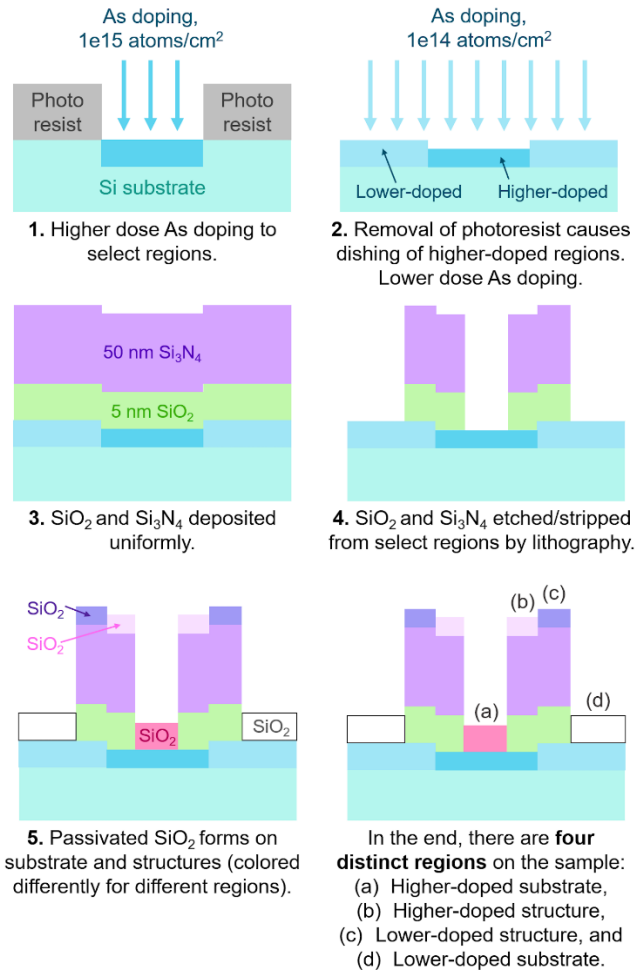

**Fig. S11: Schematic of the sample fabrication process.** Different colors correspond to different materials, and match those used in Fig. 2 of the main text.

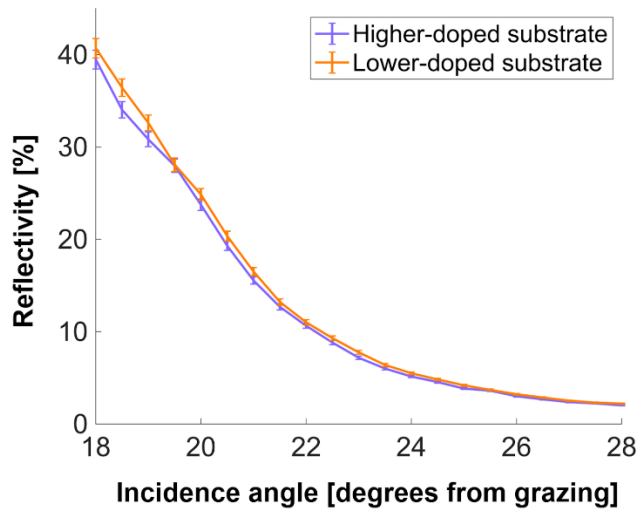

**Fig. S12: Angle-dependent reflectometry data collected from the higher-doped and lower-doped substrate.** Note that the visible difference in signal comes from the difference in both the surface roughness and the dopant concentration in the two regions.

To parametrize the As dopant level in our model, we used the As dopant curve shape measured by SIMS, but we let its scaling vary. We fixed the ratio of dopant profiles between the higher- and lower-doped regions to be the nominal value expected from the fabrication process (11x). We could have more finely parametrized the dopant curve and solved for its features – for example, the curve could have been described as a combination of an exponential near the surface followed by a half-Gaussian. However, we deemed that the number of data points, and our sensitivity to the fine features of the dopant curves, were insufficient to justify us doing so. However, we did find sensitivity to these finer parametrized features of the dopant curve in the single-parameter sensitivity test, which will be discussed later.

We also solved for the deposition rate of carbon on the surface of the sample. We assumed that the deposited thickness was linear with the illumination fluence, with some unknown slope with units of nm/hour of exposure (which should be different between the two fields of view used in the segmentation, since the beam is of different intensities in those two places). The rate was scaled at each incidence angle to account for the changing beam intensity due to the beam's projection onto the sample, as well as drop in the EUV intensity over the course of the data collection.

We could have also solved for the density of the deposited carbon, and preliminary analysis showed that we have good sensitivity to this value, however this density is fairly well known from the literature (*69*), and therefore this value was fixed at 1.2 g/cm$^3$.

Wavelength offset and incidence angle offset were solved for to take account of the discrepancy from the nominal numbers that may result from slight misalignment of our wavelength-selecting multilayer mirrors and the sample mounting stage. Ideally, this angle and wavelength offset can be fed back to the data analysis and ptychography reconstruction to increase the image fidelity.

Surface and interface roughness could have been solved for; however, AFM and HAADF-STEM measurements suggested that the surface roughness in this sample was small enough such that we did not have sufficient sensitivity. Therefore, roughness at all the interfaces were set to 0.5 nm.

**Reflectance calculation**

Before predicting reflectance of EUV light from a parameterized modeled stack, the refractive indices of each layer must be calculated for the specified wavelength. This is mostly straightforward in the EUV wavelength range, because the refractive index can be written down purely as a function of the atomic scattering factor (which is tabulated and made available by the Center for X-ray Optics) and the number density of the elements present in each layer (*14*). For the doped substrate, density was calculated assuming As was substitutionally doped (i.e. As atoms were incorporated into the structure of the Si lattice, as opposed to occupying the space between the Si lattice points).

Composition vs. depth models were discretized in 0.05 nm thick increments. The doped substrate was modeled to a depth of 60 nm, deep enough to comfortably model the nominally 30 nm deep As profile. Surface and interface roughness were modeled assuming a Gaussian distribution of roughness, meaning that the index of refraction transitions in an error function form at each interface (*59*).

Once the sample has been expressed as a stack of refractive indices, the second step is to calculate the complex reflectance from the stack. This was done using our in-house calculator written based on the Parratt's formalism (*40*, *59*). Given that the features on our samples are relatively wide, we believe that reflectance from each region are well described by reflectance calculated under the assumption of transverse homogeneity, and can use computationally cheap 0+1D methods to calculate the reflectance of the stack instead of using memory- and processor-heavy methods such as 3D rigorous coupled-wave analysis. This virtue is of immense value since it is a critical step for evaluating each of many candidates at every generation in the genetic algorithm.

**Genetic algorithm used for composition reconstruction**

A genetic algorithm was used to optimize the reconstructed sample stack. While many types of algorithms exist for optimization, genetic algorithms have been commonly used for X-ray Reflectivity (XRR) measurements due to their robustness, ability to find the global minimum when many local minima are present (*70*, *71*), and ease of implementation.

The genetic algorithm was implemented with Matlab's ga() function. The algorithm first generated candidate sample models with the parameters in each genome being tested, then calculated the phase shift upon reflection from the samples with the candidate wavelength and angle offset via the reflectance calculation described in the previous section, and then compared the predicted to the measured phase shifts. The error metric that we minimized in the genetic algorithm was the chi-square, $\chi^2$ (*54*):

$$\chi^2 = \sum_{i=1}^{N} \left(\frac{\varphi_i - \varphi(\theta_i|a)}{\sigma_i}\right)^2 \tag{S11}$$

where the summation is over $N = 10$ data points that spans over two fields of view and 5 incidence angles. $\varphi_i$ are the measured phase steps, $\varphi(\theta_i|\boldsymbol{a})$ are the calculated phase steps for the corresponding data point with incidence angle $\theta_i$ and vector of solved-for parameters $\boldsymbol{a}$, and $\sigma_i$ is the standard error of the mean for that data point.

To find the best-fit solution, the genetic algorithm was run 20 times, with 200 genomes in each population, and population generation and modification as well as function stopping settings left as Matlab's default values. Each run involved 1500 or fewer generations. The lowest-error solution from the 20 runs was then fed into a new, single run of the algorithm as the initial guess, to make sure that the found minimum is stable. The final best fit we obtained is shown in Fig. S10, and the values of the fitted parameters are reported partially in Table 1, and completely in Table S1.

**Sensitivity of the phase-sensitive imaging reflectometer to more sample parameters**

We present Supplementary Table 1, which is the complete, non-abbreviated version of Table 1 in the main text. This version of the table contains all of the parameters solved for using genetic algorithm, in particular the system calibration parameters including the C deposition rates at the two fields of view, wavelength and the incidence angle offset. In addition, we include additional parameters tested for single-parameter confidence interval, including variables in a parametrized model of As dopant concentration vs. depth curve and C density.

The atomic force microscopy (AFM) data was collected using Digital Instruments Dimension 3100, with step size of 0.16 μm for the large field of view and 0.02 μm for the zoom-in on the gratings as shown in Fig. 3F of the main text.

Energy dispersive X-ray spectroscopy (EDS) and high-angle annual dark field imaging (HAADF) were conducted twice. HAADF images in Fig. 3A and B of the main paper were taken on double aberration-corrected TEAM 1 transmission electron microscope (TEM) operated at 300 kV. The EDS image shown in Fig. 3C was collected on aberration-corrected Titan Themis S/TEM, equipped with a Super-X detector operated at 300 kV.

| | Feature | Nominal Value | Phase-Sensitive Imaging Reflectometry | | SIMS * | AFM | EDS / HAADF |
|---|---|---|---|---|---|---|---|
| | | | Simultaneous | Single-Parameter Confidence Interval | | | |
| **Layer Thickness [nm]** | SiO$_2$ on Si$_3$N$_4$ structure | 0 – 4 | (Set to 3) | ± 0.3 | — | — | 3.0 – 5.0 † |
| | Si$_3$N$_4$ | 50 | (Set to 50) | Lower bound at 30 | — | — | 41 – 45 |
| | Patterned SiO$_2$ under structure | 5 | (Set to 5) | No sensitivity at wavelength of 30 nm | — | — | 6.5 – 7.5 |
| | Structure height | — | 48.2 ± 0.2 | ± 0.02 | — | 45.0 – 45.8 | 48 – 51 |
| | SiO$_2$ on higher-doped substrate | 0 – 4 | 2.7 ± 0.3 | ± < 0.05 | — | — | 2.0 – 4.0 † |
| | SiO$_2$ on lower-doped substrate | 0 – 4 | 2.0 ± 0.3 | ± < 0.05 | — | — | 2.0 – 4.0 † |
| | Dopant-related etch depth | — | 6.09 ± 0.07 | ± 0.02 | — | 7.8 – 8.0 | 5.5 – 7.5 |
| **Interface Quality [nm]** | Average surface/interface roughness | — | (Set to 0.5) | Upper bound 0.8 | — | — | 0.5 – 1.0 |
| | Surface roughness on structures | — | (Set to 0.5) | ± 0.2 | — | 0.4 – 0.5 | — † |
| | Surface roughness on lower-doped substrate | — | (Set to 0.5) | ± 0.1 | — | 0.4 – 0.5 | — † |
| | Surface roughness on higher-doped substrate | — | (Set to 0.5) | ± 0.3 | — | 0.4 – 0.5 | — † |
| **Dopant** | Depth-integrated dose [atoms/cm$^2$] | 1.1e15 | 0.75e15 with upper bound at 5.6e15 | Upper bound at 2.1e15 | 1.05e15 | — | 1.3e15 |
| | Peak concentration [atomic %] | — | (Curve shape set by SIMS) | Upper bound at 9.3 | 3.8 | — | 3.1 – 4.1 |
| | Peak width [nm] | — | (Curve shape set by SIMS) | Upper bound at 30.3 | 2.8 | — | 5.6 |
| | Gaussian height [atomic %] | — | (Curve shape set by SIMS) | Upper bound at 3.2 | 1.1 | — | 0.8 – 1.8 |
| | Gaussian width [nm] | — | (Curve shape set by SIMS) | No sensitivity | 13.3 | — | 10.3 |
| | High/Low doped ratio | 11 | (Set to 11) | Lower bound at 0.75 | — | — | — |
| **System Calibration** | C deposition rate on wide grating FOV [nm/hr at normal incidence at initial power]‡ | — | 8.0 ± 0.6 | ± 0.05 | — | 0 – 8.3 | — |
| | C deposition rate on narrow grating FOV [nm/hr at normal incidence at initial power]‡ | — | 3.0 ± 0.1 | ± 0.3 | — | 0 – 10.6 | — |
| | C density [g/cm$^3$] | 0.8 – 1.4 | (Set to 1.2) | ± 0.01 | — | — | — |
| | Wavelength [nm] | 29.3 | 32.8 ± 0.3 | ± 0.02 | — | — | — |
| | Angle offset [deg] | 0 | 2.4 ± 0.2 | ± 0.01 | — | — | — |

\* The dopant measurements by SIMS were taken on an unpatterned sister wafer. The technique could have made similar measurements on layer thicknesses if there were wafers with the same fabrication steps as this sample, but with much bigger feature sizes.

† Variation in the SiO$_2$ thicknesses between phase-sensitive imaging reflectometry and EDS/HAADF is expected, because the sample had sufficient time to oxidize further between the two measurements. The sample was not prepared to perform surface roughness measurements.

‡ At chamber vacuum level of ~5 ×10$^{-7}$ torr.

**Table S1: Sensitivity of the phase-sensitive imaging reflectometer.** This table compares the recovered values of different sample parameters by multiple metrology techniques, and is the complete, non-abbreviated version of Table 1 in the main text. The "Nominal Value" column contains the design parameters, as well as expected values from the literature. For phase-sensitive imaging reflectometry, the "Simultaneous" column shows the values simultaneously solved for using genetic algorithm with the experimental data; only some of the sample parameters were solved for due to the limited number of data points. The "Single-Parameter" column shows the sensitivity to these parameters in a single dimension, measured by how much the fit to the data worsens if an individual parameter is varied around the found solution. This column is a rough estimate

of how low the confidence intervals could get with this dataset if we were solving for fewer parameters, and were able to fix the rest using other metrology techniques. The error bars in the Phase-Sensitive imaging reflectometry columns are given at one standard deviation, while the ranges reported for other techniques, when given, are more loosely defined reasonable ranges given to each measurement. For single-parameter confidence interval calculation, the dopant concentration vs. depth was parametrized as the concatenation of an exponential at the surface and a Gaussian extending into the bulk.